\documentclass[aps,prd,reprint,preprintnumbers,superscriptaddress,nofootinbib,longbibliography,floatfix]{revtex4-2}
\usepackage[T1]{fontenc}
\usepackage{lmodern}
\usepackage{amsmath,amssymb,bm}
\usepackage{microtype}
\usepackage[hidelinks]{hyperref}
\hypersetup{pdftitle={Energy and Charge Transmission at Conformal Interfaces: Relations and Bounds},pdfauthor={Yuma Furuta; Yuya Kusuki; Toshiki Onagi; Yin Tang}}
\newcommand{\kk}{k_{LR}}
\newcommand{\cc}{c_{LR}}

\begin{document}
\preprint{KYUSHU-HET-390}
\preprint{RIKEN-iTHEMS-Report-26}
\title{Energy and Charge Transmission at Conformal Interfaces: Relations and Bounds}
\author{Yuma Furuta}
\affiliation{\it Institute for Advanced Study, Kyushu University, Fukuoka 819-0395, Japan}
\author{Yuya Kusuki}
\affiliation{\it Institute for Advanced Study, Kyushu University, Fukuoka 819-0395, Japan}
\affiliation{\it Department of Physics, Kyushu University, Fukuoka 819-0395, Japan}
\affiliation{\it RIKEN Center for Interdisciplinary Theoretical and Mathematical Sciences (iTHEMS), Wako, Saitama 351-0198, Japan}
\affiliation{\it JST, PRESTO, 4-1-8 Honcho, Kawaguchi, Saitama 332-0012, Japan}
\author{Toshiki Onagi}
\affiliation{\it Yukawa Institute for Theoretical Physics, Kyoto University, Kitashirakawa Oiwakecho, Sakyo-ku, Kyoto 606-8502, Japan}
\author{Yin Tang}
\affiliation{\it Institute for Advanced Study, Kyushu University, Fukuoka 819-0395, Japan}
\begin{abstract}
Symmetry constrains how independently energy and charge can pass through a conformal interface.
For interfaces between Wess--Zumino--Witten (WZW) models $G_{k_L}$ and $G_{k_R}$ preserving the diagonal affine algebra, we derive an exact relation between the energy and charge transmission coefficients, $c_{LR}=2(\dim G)k_{LR}/(k_L+k_R+2h^\vee)$.
Together with positivity of outgoing energy, this relation excludes complete energy transmission at unequal levels.
With additional bulk degrees of freedom, the equality is replaced by bounds controlled by the coset central charges.
Complete transmission from a lower-level WZW model then imposes a lower bound on the coset central charge on the other side.
In linear response, the relation fixes the ratio of thermal to charge conductance to the harmonic mean of the two bulk ratios, independently of the interface, while coset sectors bound the thermal conductance at fixed charge conductance.
This interface-independent ratio is a generalized Wiedemann--Franz law for conformal interfaces between WZW models.
The charge transmission coefficient $k_{LR}$ also governs equilibrium charge fluctuations and fixes the leading quadratic dependence of the boundary-condition-changing operator's scaling dimension on the rotation angle.
\end{abstract}
\maketitle

\section{Introduction}

Conformal interfaces separate two conformal field theories (CFTs) and can transmit energy and conserved charges \cite{FFRS2004,QRW,SakaiSatoh,KRW2009,Gaiotto2012,KM2014,BrehmBrunner,BrunnerSchmidt2016,MakabeWatts,Bachas2020,Bachas2023,KKOSW2023,KKOSW2024,Furuta2026,BaigKarch2022,BaigShashi2023,BaigKarchWang2024,GutperleHultgreenMena2025,BaradTangWen2025,AntinucciEtAl2025,LiuWang2025,AbdallaChandraWang2026,ZhangEtAl2026,FurutaEtAlComplex2026,AravEtAl2026,ChenLogarithmic2026,ChenRGWalls2026,AntunesChenRong2026,Zheng2026,LiuWangZeng2025,BanerjeePolicastro2025}.
Energy and a conserved charge can be transmitted to different extents, so understanding one transport process need not determine the other.
Energy transmission is controlled by the coefficient $\cc$ of a stress-tensor two-point function across the interface, and charge transmission by the coefficient $\kk$ of a current two-point function \cite{QRW,KM2014,Billo2016,Andrei2020,MPR}.
Conformal symmetry fixes the position dependence of these correlators but does not by itself relate their coefficients.
We determine when symmetry fixes their relation and what transmission remains possible when those conditions are relaxed.

Wess--Zumino--Witten (WZW) models provide a natural setting for this question because the Sugawara construction expresses the stress tensor in terms of the currents \cite{BPZ,Witten1984,KZ1984,DiFrancesco}. Quella, Runkel, and Watts calculated energy transmission for a family of $SU(2)$ interfaces \cite{QRW}, and Kimura and Murata calculated charge transmission for the same family \cite{KM2014,KM2015}. Their explicit results imply the proportionality below for those interfaces. We derive the relation for any compact simple group $G$ and any interface between $G_{k_L}$ and $G_{k_R}$ WZW models preserving the diagonal affine $G_{k_L+k_R}$ algebra. The derivation uses only the gluing conditions and the Sugawara construction.

The resulting relation is
\begin{equation}
\cc=\frac{2\dim G}{k_L+k_R+2h^\vee}\,\kk,
\label{eq:main}
\end{equation}
where $h^\vee$ is the dual Coxeter number. The proportionality factor depends only on the bulk theories, although the transmission coefficients themselves vary with the interface. The relation also holds for automorphism-twisted gluing when $\kk$ is defined in a common basis of the preserved affine algebra.

Equation~\eqref{eq:main} fixes the ratio of the coefficients when they are nonzero. Positivity of outgoing energy also bounds their magnitudes. Using the interface scattering framework of Ref.~\cite{MPR} for states created by a current, we obtain
\begin{align}
0\le\kk&\le\frac{2[k_Lk_R+h^\vee\min(k_L,k_R)]}{k_L+k_R+2h^\vee},\label{eq:intro-k-bound}\\
0\le\cc&\le\min(c_L,c_R)\left[1-\frac{(k_L-k_R)^2}{(k_L+k_R+2h^\vee)^2}\right],
\label{eq:intro-c-bound}
\end{align}
where $c_L$ and $c_R$ are the bulk central charges. At unequal levels, $\cc$ is strictly below the smaller central charge. Thus some energy must be reflected from either side, regardless of the interface construction.

Additional bulk degrees of freedom allow energy transmission to vary at fixed charge transmission. Decomposing the stress tensor into its Sugawara and coset parts replaces Eq.~\eqref{eq:main} by upper and lower bounds controlled by the coset central charges. These bounds also quantify what is needed to restore complete transmission at unequal levels. A theory containing $SU(2)_3$ currents requires a coset central charge of at least $16/25$ to transmit all energy from $SU(2)_1$. Even without conservation of the selected charge, a quadratic lower bound relates its current correlation across the interface to energy transmission.

These constraints have a direct interpretation in thermal and charge transport. Thermal transport through an impurity can modify the Wiedemann--Franz ratio in a Luttinger liquid \cite{KaneFisher}. Plasmon reflection at the contacts of an interacting quantum wire can also modify this ratio \cite{Fazio}. Charge and energy transport have also been calculated for free-boson quantum-wire junctions \cite{GT}. Static current correlations determine charge conductance in boundary conformal field theory \cite{ChamonOA,OCA2006,Rahmani}. The thermal conductance follows from the steady-state energy current between reservoirs \cite{MPR,BernardDoyon2012,BernardDoyon2016}. For a preserved charge, the linear-response conductances at zero chemical potential are $G_Q=\kk/(2\pi)$ for a unit-norm generator and $\kappa/T=\pi\cc/6$ ($\hbar=k_B=1$), as derived in Sec.~\ref{sec:conductance}. For $G_Q>0$, Eq.~\eqref{eq:main} fixes the ratio
\begin{equation}
\frac{\kappa}{T G_Q}=\frac{2\pi^2\dim G}{3(k_L+k_R+2h^\vee)}.
\label{eq:intro-lorenz}
\end{equation}
Different interfaces can have different conductances but share this ratio. With bulk coset sectors, their central charges instead bound the thermal conductance at fixed temperature and charge conductance, including when the latter vanishes.

\section{Interface data and preserved symmetry}

\subsection{Bulk currents and folding}

We assume unitarity of the bulk CFTs and the interface throughout. Let the two bulk theories be WZW models for the same compact simple group $G$, at levels $k_L$ and $k_R$. We choose orthonormal Lie-algebra generators for which the holomorphic currents have the operator product expansion (OPE)
\begin{equation}
J_i^a(z)J_i^b(0)\sim\frac{k_i\delta^{ab}}{z^2}+\frac{i f^{ab}{}_{c}J_i^c(0)}{z},\qquad i=L,R,
\label{eq:ope}
\end{equation}
with $f^{acd}f^{bcd}=2h^\vee\delta^{ab}$. Repeated Lie-algebra indices are summed unless otherwise specified. The antiholomorphic currents obey the corresponding algebra. The full bulk stress tensors and central charges are given by the Sugawara construction,
\begin{equation}
T_i=\frac{{:J_i^aJ_i^a:}}{2(k_i+h^\vee)},\qquad c_i=\frac{k_i\dim G}{k_i+h^\vee}.
\label{eq:sugawara}
\end{equation}
We use the same orthonormal basis for the Lie algebra on the two sides.

Let $z=\tau-ix$ and $\bar z=\tau+ix$, with Euclidean time $\tau$ along the interface at $x=0$. Folding the right-hand theory turns the interface into a boundary condition for $G_{k_L}\otimes G_{k_R}$ \cite{Cardy1984,Ishibashi,CardyLewellen,AffleckLudwigG,WongAffleck,Oshikawa1996,Oshikawa1997,Bachas2002,CardyBCFT,RecknagelSchomerusBook}. Let $L_{i,n}$ and $\bar L_{i,n}$ be the modes of the full stress tensors $T_i$ and $\bar T_i$ of the folded factors. Conformal gluing of the boundary state $|B\rangle$ reads
\begin{equation}
\bigl(L_{L,n}+L_{R,n}-\bar L_{L,-n}-\bar L_{R,-n}\bigr)|B\rangle=0.
\label{eq:conformal-gluing}
\end{equation}
We further impose the diagonal affine gluing condition
\begin{equation}
\bigl(J_{L,n}^a+J_{R,n}^a+\bar J_{L,-n}^a+\bar J_{R,-n}^a\bigr)|B\rangle=0.
\label{eq:gluing}
\end{equation}
Folding the right-hand theory by $x\mapsto -x$ exchanges $J_R$ and $\bar J_R$, so current conservation becomes $J_L+J_R=\bar J_L+\bar J_R$ on the real axis. Mapping this condition to the unit circle gives the plus sign between the holomorphic and antiholomorphic modes in Eq.~\eqref{eq:gluing}, since the currents have conformal weight one. Appendix~\ref{app:kubo} relates boundary-state overlaps to local correlators. We discuss the twisted boundary conditions in Sec.~\ref{subsec:automorphism-twisted}.

Equation~\eqref{eq:gluing} equates the normal currents flowing into and out of the interface. Together with conformal gluing, it implies separate Virasoro gluing for the diagonal Sugawara tensor and its coset stress tensor in the folded picture.

\subsection{Correlators across the interface}

In the folded picture, we define $\kk$ and $\cc$ by
\begin{align}
\kk\delta^{ab}&=-\frac{\langle0|J_{L,1}^a\bar J_{R,1}^b|B\rangle}{\langle0|B\rangle},\nonumber\\
\frac{\cc}{2}&=\frac{\langle0|L_{L,2}\bar L_{R,2}|B\rangle}{\langle0|B\rangle}.
\label{eq:coeff}
\end{align}
The minus sign in the first line comes from the transformation of the weight-one current under the map to the unit circle. In the unfolded picture, the current and stress-tensor correlators across the interface are
\begin{align*}
\langle J_L^a(z)J_R^b(w)\rangle&=\frac{\kk\delta^{ab}}{(z-w)^2},\\
\langle T_L(z)T_R(w)\rangle&=\frac{\cc}{2(z-w)^4},
\end{align*}
where $z$ and $w$ are the insertion points on the left and right. Same-chirality correlators on each side retain their bulk coefficients $k_i$ and $c_i/2$, respectively. Thus an identity interface between identical theories has $\kk=k_L=k_R$ and $\cc=c_L=c_R$, while a factorizing interface has $\kk=\cc=0$.

Diagonal affine gluing and total Virasoro gluing fix the corresponding same-side overlaps,
\begin{equation}
\begin{aligned}
-\frac{\langle0|J_{i,1}^a\bar J_{i,1}^b|B\rangle}{\langle0|B\rangle}&=(k_i-\kk)\delta^{ab},\\
\frac{\langle0|L_{i,2}\bar L_{i,2}|B\rangle}{\langle0|B\rangle}&=\frac{c_i-\cc}{2},\qquad i=L,R.
\end{aligned}
\label{eq:reflection}
\end{equation}
The current reflection relation in Eq.~\eqref{eq:reflection} uses current gluing. Conformal gluing alone gives the stress-tensor relation.
The charge and energy transmission coefficients are \cite{QRW,KM2014}
\begin{equation}
\mathcal T_Q=\frac{2\kk}{k_L+k_R},\qquad \mathcal T_E=\frac{2\cc}{c_L+c_R}.
\label{eq:transmission-coefficients}
\end{equation}
Each coefficient equals one for an identity interface between identical theories. At unequal levels, the transmitted energy fractions for stress-tensor excitations incident from each side are instead $\cc/c_L$ and $\cc/c_R$ \cite{MPR}.

\section{Relation and bounds}
\label{sec:relation-and-bounds}
Diagonal current gluing relates energy and charge transmission without specifying an interface construction. Positivity of outgoing energy then bounds both coefficients and excludes complete energy transmission at unequal levels. Automorphism-twisted gluing preserves these conclusions when charge transmission is expressed in a common basis of the preserved affine algebra.

\subsection{Derivation from current gluing}

Let $\mathcal J^a=J_L^a+J_R^a$ be the diagonal current. The Sugawara expression, together with commutativity of the two WZW factors, gives the identity on the vacuum bra
\begin{equation}
\begin{aligned}
\langle0|\sum_a(J_{L,1}^a-J_{R,1}^a)\mathcal J_1^a
&=2\langle0|\bigl[(k_L+h^\vee)L_{L,2}\\
&\hspace{3em}-(k_R+h^\vee)L_{R,2}\bigr].
\end{aligned}
\label{eq:sugawara-modes}
\end{equation}
We evaluate the matrix element obtained by inserting $\bar L_{R,2}|B\rangle$ on its right. Equations~\eqref{eq:coeff} and \eqref{eq:reflection} give
\begin{multline}
\frac{\langle0|\sum_a(J_{L,1}^a-J_{R,1}^a)\mathcal J_1^a\bar L_{R,2}|B\rangle}{\langle0|B\rangle}\\
=(k_L+h^\vee)\cc-(k_R+h^\vee)(c_R-\cc).
\label{eq:stress-overlap-evaluation}
\end{multline}
For a second evaluation, use the current gluing condition and the commutator
\begin{equation*}
\mathcal J_1^a|B\rangle=-\bar{\mathcal J}_{-1}^a|B\rangle,\qquad [\bar L_{R,2},\bar{\mathcal J}_{-1}^a]=\bar J_{R,1}^a.
\end{equation*}
Moving $\bar{\mathcal J}_{-1}^a$ through $\bar L_{R,2}$ and using $\langle0|\bar{\mathcal J}_{-1}^a=0$ gives
\begin{equation}
-\frac{\langle0|(J_{L,1}^a-J_{R,1}^a)\bar J_{R,1}^a|B\rangle}{\langle0|B\rangle}=(2\kk-k_R)\dim G.
\label{eq:current-overlap-evaluation}
\end{equation}
The last equality uses the current overlaps in Eqs.~\eqref{eq:coeff} and \eqref{eq:reflection}. Equating Eqs.~\eqref{eq:stress-overlap-evaluation} and \eqref{eq:current-overlap-evaluation}, then using $(k_R+h^\vee)c_R=(\dim G)k_R$, gives
\begin{equation}
(k_L+k_R+2h^\vee)\cc=2(\dim G)\kk.
\label{eq:relation-proof-final}
\end{equation}
This proves Eq.~\eqref{eq:main} without constructing $|B\rangle$ explicitly.

\subsection{Bounds on \texorpdfstring{$k_{LR}$ and $c_{LR}$}{kLR and cLR}}

Positivity of outgoing energy strengthens the bounds from unitarity and excludes complete energy transmission at unequal levels.
We first obtain the unitarity bounds.

The regions $x<0$ and $x>0$ are occupied by the left and right CFTs, respectively. To distinguish incoming and outgoing excitations, we analytically continue Euclidean time to $\tau=it$, yielding $z=i(t-x)$ and $\bar z=i(t+x)$. A holomorphic current $J_i$ moves to the right and an antiholomorphic current $\bar J_i$ moves to the left. For a fixed Lie-algebra index $a$, we define incoming and outgoing states of unit norm with a common Euclidean time regulator $\epsilon>0$ \cite{MPR},
\begin{equation}
\begin{aligned}
|\mathrm{in},L,a\rangle&=\frac{2\epsilon}{\sqrt{k_L}}J_L^a(-\epsilon)|0\rangle_{\!\mathcal I},\\
|\mathrm{in},R,a\rangle&=\frac{2\epsilon}{\sqrt{k_R}}\bar J_R^a(-\epsilon)|0\rangle_{\!\mathcal I},\\
|\mathrm{out},L,a\rangle&=\frac{2\epsilon}{\sqrt{k_L}}\bar J_L^a(-\epsilon)|0\rangle_{\!\mathcal I},\\
|\mathrm{out},R,a\rangle&=\frac{2\epsilon}{\sqrt{k_R}}J_R^a(-\epsilon)|0\rangle_{\!\mathcal I},
\end{aligned}
\label{eq:in-out-states}
\end{equation}
where $|0\rangle_{\!\mathcal I}$ is the vacuum in the presence of the interface. With this choice the wave packets meet the interface at $x=0$ and $t=0$. The two incoming states are orthogonal, as are the two outgoing states, because the corresponding currents belong to different folded factors and have the same chirality. The overlaps $S_{ij}=\langle\mathrm{out},i,a|\mathrm{in},j,a\rangle$ follow from Eqs.~\eqref{eq:coeff} and~\eqref{eq:reflection} and are given by
\begin{equation}
S=\begin{pmatrix}1-\kk/k_L&\kk/\sqrt{k_Lk_R}\\ \kk/\sqrt{k_Lk_R}&1-\kk/k_R\end{pmatrix}.
\label{eq:current-overlap}
\end{equation}
Reflection positivity of the Gram matrix of the four states implies $S^\dagger S\le\mathbf 1$. From Eq.~\eqref{eq:current-overlap}, the eigenvalues of $S^\dagger S$ are
\begin{equation}
1,\qquad \left(1-\frac{(k_L+k_R)\kk}{k_Lk_R}\right)^2.
\end{equation}
Together with Eq.~\eqref{eq:main}, this implies
\begin{align}
0\le\kk&\le\frac{2k_Lk_R}{k_L+k_R},\nonumber\\
0\le\cc&\le\frac{4(\dim G)k_Lk_R}{(k_L+k_R)(k_L+k_R+2h^\vee)}.
\label{eq:unitarity-bounds}
\end{align}
The coefficient $\kk$ can exceed the smaller bulk level, as an example below shows.

We now use positivity of outgoing averaged null energy in the interface scattering setup of Ref.~\cite{MPR} for states created by a current. The amplitude for an incoming current state on side $i$ to emerge as a current state on the opposite side is $\kk/\sqrt{k_Lk_R}$. The current OPE fixes its total transmitted energy fraction to $\cc/c_i$ when the full stress tensor is given by the Sugawara construction. Appendix~\ref{app:bound} gives the derivation. The energy in this one outgoing channel cannot exceed the total, so incidence from either side gives
\begin{equation}
\frac{\kk^2}{k_Lk_R}\le\min\!\left(\frac{\cc}{c_L},\frac{\cc}{c_R}\right).
\label{eq:energy-current-bound}
\end{equation}
Substituting Eq.~\eqref{eq:main} yields
\begin{align}
0\le\kk&\le\frac{2[k_Lk_R+h^\vee\min(k_L,k_R)]}{k_L+k_R+2h^\vee},\label{eq:k-strong-bound}\\
0\le\cc&\le\min(c_L,c_R)\left[1-\frac{(k_L-k_R)^2}{(k_L+k_R+2h^\vee)^2}\right].\label{eq:c-strong-bound}
\end{align}
For $k_L=k_R$, these reduce to $\kk\le k_L$ and $\cc\le c_L$, saturated by the identity interface. At unequal levels the upper bound on $\cc$ is strictly below the smaller bulk central charge.

\subsection{Automorphism-twisted gluing}
\label{subsec:automorphism-twisted}
Automorphism-twisted gluing changes the identification of charge components across the interface. The relation and bounds retain the same form when $\kk$ is defined in a common basis of the preserved diagonal affine algebra.
In the unfolded picture, the gluing conditions take the form
\begin{equation}
R_L(J_L-\Omega_L\bar J_L)=R_R(J_R-\Omega_R\bar J_R),
\label{eq:twisted-gluing}
\end{equation}
where $R_i$ and $\Omega_i$ are orthogonal Lie-algebra automorphisms implemented in the corresponding WZW theories \cite{AlekseevSchomerus,BPPZ,MPR}. In the folded picture, the two chiralities on the right are exchanged. In a common basis of the preserved affine algebra, the holomorphic currents are $R_LJ_L$ and $R_R\Omega_RJ_R$, and the antiholomorphic ones are $R_L\Omega_L\bar J_L$ and $R_R\bar J_R$. They obey Eq.~\eqref{eq:gluing} and have unchanged Sugawara tensors, so Eq.~\eqref{eq:main} and the bounds hold for $\kk$ in this basis.

To express current correlators in the original basis, define the matrix of coefficients across the interface by
\begin{equation}
M^{ab}=-\frac{\langle0|J_{L,1}^a\bar J_{R,1}^b|B\rangle}{\langle0|B\rangle}.
\label{eq:twist-matrix-def}
\end{equation}
The two currents in this correlator transform by $R_L$ and $R_R$, respectively. Consequently, for arbitrary $\Omega_L$ and $\Omega_R$,
\begin{equation}
\begin{gathered}
R_LMR_R^{\mathsf T}=\kk\mathbf1,\qquad M=\kk R_L^{\mathsf T}R_R,\\
\cc=\frac{2\dim G}{k_L+k_R+2h^\vee}\kk.
\end{gathered}
\label{eq:twist-matrix}
\end{equation}
Here $\kk=(\dim G)^{-1}\operatorname{tr}(R_LMR_R^{\mathsf T})$. The $\Omega_i$ do not appear explicitly in $M$ because it pairs $J_L$ with $\bar J_R$ in the folded picture, but they enter the opposite pairing,
\begin{equation}
-\frac{\langle0|J_{R,1}^a\bar J_{L,1}^b|B\rangle}{\langle0|B\rangle}=\kk(\Omega_R^{\mathsf T}R_R^{\mathsf T}R_L\Omega_L)^{ab}.
\label{eq:twist-opposite-pairing}
\end{equation}
With $R_L=R_R$, $M=\kk\mathbf1$. Rotating one side of an identity $SU(2)_k$ interface by $\pi$ around the first axis instead gives $M=\operatorname{diag}(k,-k,-k)$, with $\kk=k$ in the common basis and $\cc=3k/(k+2)$. Thus neither an entry of $M$ nor its trace in the original basis generally replaces $\kk$ in Eq.~\eqref{eq:main}. The same argument applies to outer automorphisms implemented in the WZW theory.

\section{Transmission bounds under weaker assumptions}\label{sec:weaker}

When the bulk stress tensors include coset contributions, energy transmission can vary at fixed charge transmission, but the coset central charges bound the allowed range. These bounds also constrain the additional degrees of freedom required for complete transmission at unequal levels. We first retain diagonal current gluing and determine these constraints, then show that a quadratic lower bound on energy transmission survives even without conservation of the selected charge.

\subsection{Preserved affine symmetry with coset stress tensors}\label{sec:coset-bound}

Suppose each bulk theory contains the affine algebra $G_{k_i}$ and the interface preserves the diagonal currents. Write $T_i=T_i^G+T_i^\perp$, where $T_i^G$ is the Sugawara stress tensor and $T_i^\perp$ is the coset stress tensor, with central charges $c_i^G=k_i\dim G/(k_i+h^\vee)$ and $c_i^\perp=c_i-c_i^G$. Define $b_i$ by
\begin{equation}
b_i=\frac{2\langle0|L_{i,2}^\perp\bar L_{j,2}|B\rangle}{\langle0|B\rangle},\qquad j\ne i,
\label{eq:coset-mixed}
\end{equation}
where $L_{i,n}^\perp$ denotes a mode of $T_i^\perp$. Repeat the preceding calculation with the Sugawara modes in Eq.~\eqref{eq:sugawara-modes} and the full stress tensor on the outgoing side. The evaluation using the current commutators is unchanged, and the difference is the contribution from $T_i^\perp$, giving
\begin{equation}
\begin{split}
(k_L+k_R+2h^\vee)\cc-2(\dim G)\kk\\
=(k_L+h^\vee)b_L+(k_R+h^\vee)b_R.
\end{split}
\label{eq:coset-gluing-relation}
\end{equation}

Positivity of the outgoing energy in each sector bounds the coset contributions. Let $\mathcal T$ be the Sugawara stress tensor of the diagonal current $\mathcal J^a$, of level $k_L+k_R$ and central charge $c_{\mathcal J}$. Let $T_0$ be the stress tensor of the diagonal coset $(G_{k_L}\times G_{k_R})/G_{k_L+k_R}$, with central charge $c_0$. Then
\begin{equation}
\begin{aligned}
c_{\mathcal J}&=\frac{(k_L+k_R)\dim G}{k_L+k_R+h^\vee},\\
T_0&=T_L^G+T_R^G-\mathcal T,\\
c_0&=c_L^G+c_R^G-c_{\mathcal J}\\
&=\frac{(\dim G)k_Lk_R(k_L+k_R+2h^\vee)}{(k_L+h^\vee)(k_R+h^\vee)(k_L+k_R+h^\vee)}.
\end{aligned}
\label{eq:T0-def}
\end{equation}
Both $\mathcal T$ and $T_L+T_R-\mathcal T=T_0+T_L^\perp+T_R^\perp$ obey Virasoro gluing, by Eqs.~\eqref{eq:conformal-gluing} and \eqref{eq:gluing}. For weight-two holomorphic fields $A,A'$, write $(A,A')$ for the coefficient of $(z-w)^{-4}$ in their vacuum two-point function and
\begin{equation}
\mathcal R(A,A')=\frac{\langle0|A_2\bar A'_2|B\rangle}{\langle0|B\rangle}
\end{equation}
for the overlap in the folded picture, so that $\cc/2=\mathcal R(T_L,T_R)$ and $b_L/2=\mathcal R(T_L^\perp,T_R)$. Virasoro gluing gives $\mathcal R(A,\mathcal T)=(A,\mathcal T)$ and $\mathcal R(A,T_L+T_R-\mathcal T)=(A,T_L+T_R-\mathcal T)$ for any $A$, and likewise with the two arguments exchanged.

The Sugawara construction gives
\[
(T_L^G,\mathcal T)=\frac{k_L}{k_L+k_R}\frac{c_{\mathcal J}}{2},
\qquad
(T_L^G,T_0)=\mu_L\frac{c_0}{2},
\]
where
\begin{equation}
\mu_L=\frac{k_R+h^\vee}{k_L+k_R+2h^\vee},\qquad
\mu_R=\frac{k_L+h^\vee}{k_L+k_R+2h^\vee},
\label{eq:mu-def}
\end{equation}
and $\mu_L+\mu_R=1$. Hence
\begin{equation}
\begin{aligned}
T_L^G&=\frac{k_L}{k_L+k_R}\mathcal T+\mu_LT_0+V,\\
T_R^G&=\frac{k_R}{k_L+k_R}\mathcal T+\mu_RT_0-V,\\
V&=\frac{\sum_a:\mathcal J^a\mathcal K^a:}{(k_L+k_R)(k_L+k_R+2h^\vee)},
\end{aligned}
\label{eq:TG-decomp}
\end{equation}
where $\mathcal K^a=k_RJ_L^a-k_LJ_R^a$ is orthogonal to $\mathcal J^a$ in the vacuum two-point function, and $V$ is orthogonal to $\mathcal T$ and $T_0$. If a field $A$ has a regular OPE with $\mathcal J^a$, then $\mathcal R(A,\mathcal T)=(A,\mathcal T)=0$, and $\mathcal R(A,V)=0$ by current gluing. Writing
\[
\langle0|\bar V_2\propto\sum_a\langle0|\bar{\mathcal K}_1^a\bar{\mathcal J}_1^a
\]
and using $\bar{\mathcal J}_1^a|B\rangle=-\mathcal J_{-1}^a|B\rangle$, we see that the mode $\mathcal J_{-1}^a$ commutes with $A_2$ and $\bar{\mathcal K}_1^a$ and annihilates $\langle0|$. Since $T_L^\perp$, $T_R^\perp$ and $T_0$ have regular OPEs with $\mathcal J^a$,
\begin{equation}
\mathcal R(A,T_i^G)=\mu_i\mathcal R(A,T_0),\qquad
A\in\{T_L^\perp,T_R^\perp,T_0\}.
\label{eq:G-sector}
\end{equation}
In particular,
\begin{align*}
b_L&=2\mu_R\mathcal R(T_L^\perp,T_0)+2\mathcal R(T_L^\perp,T_R^\perp),\\
b_R&=2\mu_L\mathcal R(T_R^\perp,T_0)+2\mathcal R(T_R^\perp,T_L^\perp),
\end{align*}
and Eq.~\eqref{eq:coset-gluing-relation} reads
\begin{equation}
\begin{aligned}
\Delta&\equiv\cc-\frac{2\dim G}{k_L+k_R+2h^\vee}\kk=\mu_Rb_L+\mu_Lb_R\\
&=2\mu_R^2\mathcal R(T_L^\perp,T_0)+2\mu_L^2\mathcal R(T_R^\perp,T_0)\\
&\quad+2\mu_R\mathcal R(T_L^\perp,T_R^\perp)+2\mu_L\mathcal R(T_R^\perp,T_L^\perp).
\end{aligned}
\label{eq:Delta-def}
\end{equation}

The averaged null energy condition (ANEC) applies to the full outgoing stress tensors and to each of $\bar T_L^G$, $\bar T_L^\perp$, $\bar T_R^G$, $\bar T_R^\perp$ and $\bar T_0$. Each generates a Virasoro algebra that acts unitarily on the outgoing states, with $L_0$ bounded below by zero because the bulk modules decompose into unitary affine and coset modules \cite{GKO1985,GKO1986}. The averaged null energy operator of such a stress tensor is nonnegative on every state \cite{HofmanMaldacena,FLPW,HKT,MPR}. Hence the outgoing energy in each of these sectors is separately nonnegative on scattering states. For a state created by a field $A$ whose OPE with itself contains $2A(0)/z^2$ as its only weight-two term, the three-point function with the outgoing stress tensor $A'$ of a sector is fixed by this term, and the outgoing energy in that sector is proportional to the corresponding overlap $\mathcal R(A,A')$ with a positive coefficient, as in Appendix~\ref{app:bound}. This applies to $A\in\{T_L^\perp,T_R^\perp,T_0\}$, which generate commuting Virasoro algebras. Therefore all overlaps satisfy
\begin{equation}
\mathcal R(A,A')\ge0,\qquad A,A'\in\{T_L^\perp,T_R^\perp,T_0\},
\label{eq:sector-positivity}
\end{equation}
and Virasoro gluing of $T_0+T_L^\perp+T_R^\perp$ fixes their row and column sums,
\begin{equation}
\sum_{A'}\mathcal R(A,A')=(A,A),\qquad
\sum_{A}\mathcal R(A,A')=(A',A'),
\label{eq:sector-sums}
\end{equation}
with $(T_i^\perp,T_i^\perp)=c_i^\perp/2$ and $(T_0,T_0)=c_0/2$. Dropping nonnegative terms from Eq.~\eqref{eq:sector-sums} gives
\begin{align*}
\mathcal R(T_L^\perp,T_0)+\mathcal R(T_L^\perp,T_R^\perp)&\le c_L^\perp/2,\\
\mathcal R(T_R^\perp,T_0)+\mathcal R(T_R^\perp,T_L^\perp)&\le c_R^\perp/2,\\
\mathcal R(T_L^\perp,T_R^\perp)&\le c_R^\perp/2,\\
\mathcal R(T_L^\perp,T_0)+\mathcal R(T_R^\perp,T_0)&\le c_0/2.
\end{align*}
Take $c_L^\perp\ge c_R^\perp$. Writing $\mu_R=\mu_R^2+\mu_L\mu_R$ and $\mu_L=\mu_L^2+\mu_L\mu_R$ in the last two terms of Eq.~\eqref{eq:Delta-def}, the first three inequalities give
\begin{equation*}
\Delta\le\mu_R^2c_L^\perp+\mu_L^2c_R^\perp+2\mu_L\mu_Rc_R^\perp,
\end{equation*}
and the last three give 
\begin{equation*}
\Delta\le\mu_R^2c_0+c_R^\perp.
\end{equation*}
Together with $\Delta=\mu_Rb_L+\mu_Lb_R\ge0$, this gives
\begin{equation}
\begin{aligned}
\frac{2\dim G}{k_L+k_R+2h^\vee}\kk&\le\cc,\\
\cc&\le\frac{2\dim G}{k_L+k_R+2h^\vee}\kk\\
&\quad+c_R^\perp+\mu_R^2\min\bigl(c_L^\perp-c_R^\perp,\,c_0\bigr),
\end{aligned}
\label{eq:coset-window}
\end{equation}
for $c_L^\perp\ge c_R^\perp$, and with $L\leftrightarrow R$ otherwise. The energy transmitted in excess of the amount fixed by charge transmission is bounded by the smaller coset central charge, plus $\mu^2$ times the smaller of the mismatch and $c_0$, where $\mu$ is the weight of the side with the smaller coset. The weaker bound obtained from $0\le b_i\le c_i^\perp$ alone, $\Delta\le\mu_Rc_L^\perp+\mu_Lc_R^\perp$, is improved whenever $c_L^\perp\ne c_R^\perp$. The general energy bound $\cc\le\min(c_L,c_R)$ also applies. For WZW bulk theories $c_i^\perp=0$, the bounds coincide and recover Eq.~\eqref{eq:main}.

At equal levels $k_L=k_R=k$ and equal central charges $c_L=c_R=c$, the overlap matrix in Eq.~\eqref{eq:current-overlap} gives $0\le\kk/k\le1$. Since $\mu_L=\mu_R=1/2$ and $c_L^\perp=c_R^\perp=c-c^G$, Eq.~\eqref{eq:coset-window} simplifies to
\begin{equation}
c^G\frac{\kk}{k}\le\cc\le c^G\frac{\kk}{k}+c-c^G,\qquad c^G=\frac{k\dim G}{k+h^\vee}.
\label{eq:g-window}
\end{equation}
In particular, complete reflection of the preserved currents, $\kk=0$, implies $\cc\le c-c^G$, whereas complete energy transmission, $\cc=c$, requires $\kk=k$.

The restriction does not require preservation of the full non-Abelian symmetry. For a single preserved $U(1)$ current at equal levels $k$ and bulk central charges $c$, its Sugawara stress tensor $T_i^J={:J_iJ_i:}/(2k_i)$ has central charge one, while $T_i^\perp=T_i-T_i^J$ has central charge $c-1$. Repeating the calculation for this current gives
\begin{equation}
0\le\mathcal T_Q=\frac{\kk}{k}\le1,\qquad \mathcal T_Q\le\cc\le\mathcal T_Q+c-1.
\label{eq:u1-window}
\end{equation}
For an interface between two $SU(2)_k$ WZW models that preserves a Cartan current, $c=3k/(k+2)$. Complete reflection of this charge therefore implies
\begin{equation*}
\mathcal T_E=\frac{\cc}{c}\le\frac{2(k-1)}{3k}.
\end{equation*}
At level two, $\mathcal T_E\le1/3$.

The realization of $SU(2)_2$ in terms of three free Majorana fermions \cite{Oliviero2024} gives interfaces that saturate both bounds on $\cc$ in Eq.~\eqref{eq:u1-window}. With $\chi_a(z)\chi_b(0)\sim\delta_{ab}/z$, the Cartan current in the normalization of Eq.~\eqref{eq:ope} is $J^3=\sqrt{2}\,i\chi_1\chi_2$. Choose identical scattering matrices for $\chi_1$ and $\chi_2$, with energy transmission coefficient $\mathcal T_1$. Marginal interface couplings let $\mathcal T_1$ and the energy transmission coefficient $\mathcal T_3$ of $\chi_3$ vary independently between zero and one \cite{BachasIsing2013,MPR}. Wick contractions give
\begin{equation}
\mathcal T_Q=\mathcal T_1,\qquad \cc=\mathcal T_1+\frac{\mathcal T_3}{2}.
\label{eq:majorana-window}
\end{equation}
Varying $\mathcal T_3$ saturates both bounds at fixed $\mathcal T_Q=\mathcal T_1$. At $(\mathcal T_1,\mathcal T_3)=(0,1)$, $\chi_3$ carries the maximum allowed energy, $\mathcal T_E=1/3$, while the charge is fully reflected. Thus the range in Eq.~\eqref{eq:u1-window} is sharp for this interface family.

\subsection{Coset central charge required for complete transmission}

At unequal levels, the WZW bound \eqref{eq:c-strong-bound} excludes complete energy transmission. Additional bulk degrees of freedom can remove this obstruction, but the required coset central charge has a lower bound. Let $a_i$ be the transmitted energy fraction for an incoming current state on side $i$. The coefficient $b_i$ is defined in Eq.~\eqref{eq:coset-mixed}. Current gluing and the stress-tensor decomposition give
\begin{equation}
\begin{aligned}
k_La_L+k_Ra_R&=2\kk,\\
\cc&=c_L^Ga_L+b_L=c_R^Ga_R+b_R,\\
0&\le b_i\le c_i^\perp.
\end{aligned}
\label{eq:current-coset-relation}
\end{equation}
The sector bounds of Sec.~\ref{sec:coset-bound} sharpen the last line when one side is a WZW model. If $c_L^\perp=0$, then $T_L^\perp=0$ and $b_R=2\mu_L\mathcal R(T_R^\perp,T_0)$. A right coset excitation can transmit energy only into the left $G$ sector. Equations~\eqref{eq:sector-positivity} and \eqref{eq:sector-sums} then give
\begin{equation}
0\le b_R\le\mu_L\min\bigl(c_R^\perp,\,c_0\bigr).
\label{eq:one-sided-coset-bound}
\end{equation}

Consider a $G_{k_L}$ WZW model on the left, $c_L=c_L^G$ and $c_L^\perp=0$, joined to a theory containing $G_{k_R}$ with $k_R>k_L$. Suppose $\cc=c_L$. Since the left bulk is WZW, its incoming current state has $a_L=1$. The amplitude for a left current to return in the same current channel is $1-\kk/k_L$. Its squared modulus is bounded by the reflected energy fraction $1-\cc/c_L=0$. Hence $\kk=k_L$, and Eq.~\eqref{eq:current-coset-relation} gives $a_R=k_L/k_R$ and $b_R=c_L^G-c_R^Gk_L/k_R$. The condition $b_R\le\mu_Lc_0$ in Eq.~\eqref{eq:one-sided-coset-bound} holds for all $k_R>k_L$, since
\[
\frac{b_R}{\mu_Lc_0}=\frac{(k_R-k_L)(k_L+k_R+h^\vee)}{k_R(k_R+h^\vee)}\le1.
\]
The condition $b_R\le\mu_Lc_R^\perp$ requires
\begin{equation}
\begin{aligned}
c_R^\perp&\ge\frac{b_R}{\mu_L}=\frac{1}{\mu_L}\Bigl(c_L^G-c_R^G\frac{k_L}{k_R}\Bigr)\\
&=\frac{(\dim G)\,k_L(k_R-k_L)\,(k_L+k_R+2h^\vee)}{(k_L+h^\vee)(k_R+h^\vee)^2}.
\end{aligned}
\label{eq:coset-cost}
\end{equation}
This condition can rule out complete transmission using only the bulk current levels and central charges, without constructing an interface. The factor $1/\mu_L>1$ strengthens the condition $c_R^\perp\ge b_R$ that follows from $b_R\le c_R^\perp$ alone.

For $SU(2)_1$ on the left and an $SU(2)_3$ current algebra on the right, $b_R=2/5$ and $\mu_L=5/8$, so Eq.~\eqref{eq:coset-cost} requires $c_R^\perp\ge16/25$. Since $c_R^G=9/5$, no interface preserving the diagonal $SU(2)$ currents can transmit all energy from the left if $c_R<61/25$, although the general bound $\cc\le\min(c_L,c_R)$ permits it.

The condition can also exclude a theory that has additional degrees of freedom. For $SU(2)_5\times\mathrm{Ising}$ on the right, $b_R=4/7$ and $\mu_L=7/10$, so the required coset central charge is at least $40/49$, but the Ising factor contributes only $c_R^\perp=1/2$. Hence complete transmission from $SU(2)_1$ is impossible under diagonal current gluing, even though $c_R=37/14>c_L=1$.

Conversely, additional degrees of freedom can restore complete transmission. For $SU(2)_1\times SU(2)_2$ on the right, transmit the shared $SU(2)_1$ factor and reflect the $SU(2)_2$ factor, with the diagonal current of the right-hand factors at level three. This interface has $\cc=\kk=1$, $c_R=5/2$, and $c_R^\perp=7/10$, which exceeds the required $16/25$ by $3/50$. Equation~\eqref{eq:coset-cost} is nearly saturated.

The coset central charge also limits how close an interface can come to complete transmission. For the left WZW model, positivity of reflected energy for an incoming current gives
\begin{equation}
1-\frac{\cc}{c_L}\ge
\left|1-\frac{\kk}{k_L}\right|^2.
\label{eq:reflected-current-cost}
\end{equation}
Combining this inequality with Eq.~\eqref{eq:coset-gluing-relation}, $b_L=0$, and Eq.~\eqref{eq:one-sided-coset-bound} gives
\begin{align}
&(k_L+k_R+2h^\vee)\cc\le 2k_L\bigl(1+\sqrt{1-\cc/c_L}\bigr)\dim G\nonumber\\
&\qquad+\frac{(k_R+h^\vee)^2}{k_L+k_R+2h^\vee}\min\bigl(c_R^\perp,\,c_0\bigr).
\label{eq:near-perfect-cost}
\end{align}
At $\cc=c_L$ this reduces to Eq.~\eqref{eq:coset-cost}. Setting $c_R^\perp=0$ recovers the WZW upper bound \eqref{eq:c-strong-bound}.

\subsection{A quadratic bound without assuming charge conservation}

Even if the interface does not preserve a selected charge, a nonzero current correlation across it requires nonzero energy transmission. Consider a holomorphic $U(1)$ current $J_i$ of positive level $k_i$ in each bulk theory, with $J_i(z)J_i(0)\sim k_i/z^2$. A Cartan current is also sufficient. Chirality and conformal invariance fix the correlator across the interface at separated points to
\begin{equation*}
\langle J_L(z)J_R(w)\rangle=\frac{\kk}{(z-w)^2}.
\end{equation*}
Thus $\kk$ is defined even when the interface does not preserve the charge.

Without current gluing, the reflected current correlators are not fixed by $\kk$, which need not determine a charge conductance. The bulk decomposition $T_i=T_i^J+T_i^\perp$ still holds, with central charges $1$ and $c_i-1$.

For incidence from the left, decompose $\cc=a_L+b_L$ into the contributions of $T_L^J$ and $T_L^\perp$. The current OPE identifies $a_L$ as the transmitted energy fraction for a state created by $J_L$, whose overlap with the outgoing current state is $\kk/\sqrt{k_Lk_R}$. Positivity of outgoing energy gives
\begin{equation}
a_L\ge\frac{|\kk|^2}{k_Lk_R},\qquad b_L\ge0,
\qquad\Longrightarrow\qquad
\cc\ge\frac{|\kk|^2}{k_Lk_R}.
\label{eq:quadratic-lower}
\end{equation}
For incidence from the right, the analogous bound involves the antiholomorphic current correlator across the interface.

The bound can be saturated. A conformal interface between free bosons at equal current level $k$ can have $\kk/k=t$ and $\cc=t^2$ with $t>0$ arbitrarily close to zero \cite{BachasBrunner2008,BachasWorldsheet2012,GT,MPR}. Thus no universal positive linear lower bound on $\cc$ in terms of $|\kk|/k$ survives without charge conservation. In particular, Eq.~\eqref{eq:quadratic-lower} implies that $\cc=0$ forces $\kk=0$.

\section{Unequal levels and coset examples}

For WZW bulk theories with diagonal current gluing, the energy and charge transmission coefficients coincide at equal levels. At unequal levels, Eq.~\eqref{eq:main} makes the energy transmission coefficient larger whenever transmission is nonzero. The coset construction gives explicit interfaces that illustrate both cases and shows that $\kk$ can exceed the smaller bulk current level.

\subsection{The effect of unequal levels}

Let $K=k_L+k_R$ be the level of the diagonal affine algebra. When $\kk\ne0$, substituting Eq.~\eqref{eq:main} and the Sugawara central charges into Eq.~\eqref{eq:transmission-coefficients} gives
\begin{align}
\frac{\mathcal T_E}{\mathcal T_Q}&=\frac{2K(k_L+h^\vee)(k_R+h^\vee)}{(K+2h^\vee)(2k_Lk_R+h^\vee K)}\nonumber\\
&=1+\frac{h^\vee(k_L-k_R)^2}{(K+2h^\vee)(2k_Lk_R+h^\vee K)}.
\label{eq:transmission-ratio}
\end{align}
At equal levels the coefficients coincide. At unequal levels, energy has the larger transmission coefficient for any interface with $\kk>0$.

\subsection{Interfaces from the coset construction}\label{sec:construction}

Consider interfaces constructed in the folded picture from the coset of $G_{k_L}\otimes G_{k_R}$ with respect to the diagonal $G_K$ algebra \cite{GKO1985,GKO1986,QS}. Recall the current $\mathcal K^a=k_RJ_L^a-k_LJ_R^a$, orthogonal to $\mathcal J^a$ in the vacuum two-point function.
Its boundary-state overlap defines $\omega$ by
\begin{equation}
\frac{\langle0|\mathcal K_1^a\bar{\mathcal K}_1^b|B\rangle}{\langle0|B\rangle}=-\omega k_Lk_RK\delta^{ab}.
\label{eq:omega-def}
\end{equation}
With
\begin{equation*}
J_L=\frac{k_L\mathcal J+\mathcal K}{K},\qquad J_R=\frac{k_R\mathcal J-\mathcal K}{K},
\end{equation*}
the current overlap and Eq.~\eqref{eq:main} become
\begin{align}
\kk&=\frac{k_Lk_R}{K}(1-\omega),\label{eq:komega}\\
\cc&=\frac{2(\dim G)k_Lk_R}{K(K+2h^\vee)}(1-\omega).\label{eq:comega}
\end{align}

The boundary states of Ref.~\cite{QS} are built from a factorizing interface. Let $|B_0\rangle=|\rho_L\rangle\otimes|\rho_R\rangle$ be a product of Cardy states of $G_{k_L}$ and $G_{k_R}$ \cite{Cardy1989,BPPZ,FRS2002}.
It obeys $(J^a_{i,n}+\bar J^a_{i,-n})|B_0\rangle=0$ for $i=L$ and $i=R$ separately, so it preserves the diagonal $G_K$ algebra and the coset algebra.
For an integrable representation $\rho$ of $G_K$, the states of Ref.~\cite{QS} are $|B_\rho\rangle=\mathcal L_\rho|B_0\rangle$, where $\mathcal L_\rho$ multiplies the component of a boundary state in which $G_K$ is in the representation $\nu$ by $S^{(K)}_{\rho\nu}/S^{(K)}_{0\nu}$ \cite{PetkovaZuber,GrahamWatts,BachasGaberdiel,FFRS2007}.
Here $S^{(K)}$ is the modular $S$-matrix of $G_K$ \cite{Verlinde1988}. For $|B_0\rangle$ itself, $\bar{\mathcal K}^a_1|B_0\rangle=-\mathcal K^a_{-1}|B_0\rangle$, and Eq.~\eqref{eq:omega-def} gives $\omega=1$.
The state $\mathcal K^a_{-1}|0\rangle$ is a $G_K$ primary in the adjoint representation, while $|0\rangle$ is in the vacuum representation.
Hence
\begin{equation}
\omega_\rho=\frac{S^{(K)}_{00}S^{(K)}_{\rho,{\rm adj}}}{S^{(K)}_{\rho0}S^{(K)}_{0,{\rm adj}}}.
\label{eq:generalomega}
\end{equation}
The labels $\rho_L$ and $\rho_R$ drop out, and $\rho=0$ gives the factorizing interface.

\subsection{\texorpdfstring{$SU(2)$}{SU(2)} examples}

The known $SU(2)$ coset interfaces illustrate Eq.~\eqref{eq:main} at different transmission coefficients \cite{QRW,KM2014}. With $\dim SU(2)=3$ and $h^\vee=2$, it reads
\begin{equation}
\cc=\frac{6}{k_L+k_R+4}\,\kk.
\label{eq:su2}
\end{equation}
For an interface between $SU(2)_1$ and $SU(2)_3$, the diagonal algebra is $SU(2)_4$. Inserting
\begin{equation*}
S^{(4)}_{jj'}=\frac{1}{\sqrt{3}}\sin\!\left[\frac{(2j+1)(2j'+1)\pi}{6}\right]
\end{equation*}
into Eq.~\eqref{eq:generalomega} gives $\omega_\rho=1,0,-1/2$ for $\rho=0,1/2,1$, respectively \cite{KM2014}. Equations~\eqref{eq:komega} and \eqref{eq:comega} then give Table~\ref{tab:su2}.

\begin{table}[t]
\centering
\setlength{\tabcolsep}{4pt}
\begin{tabular}{c|cc|cc}
$\rho$ & $\kk$ & $\cc$ & $\mathcal T_Q$ & $\mathcal T_E$\\\hline
$0$ & $0$ & $0$ & $0$ & $0$\\
$\frac12$ & $\frac34$ & $\frac9{16}$ & $\frac38$ & $\frac{45}{112}$\\
$1$ & $\frac98$ & $\frac{27}{32}$ & $\frac9{16}$ & $\frac{135}{224}$
\end{tabular}
\caption{The coefficients $k_{LR}$ and $c_{LR}$ and the transmission coefficients $\mathcal T_Q$ and $\mathcal T_E$ for three interfaces between $SU(2)_1$ and $SU(2)_3$ from the coset construction. The label $\rho$ is the spin of the diagonal $SU(2)_4$ representation. The bulk central charges are $c_L=1$ and $c_R=9/5$.}
\label{tab:su2}
\end{table}

Both transmitting rows obey $\cc/\kk=3/4$ and $\mathcal T_E/\mathcal T_Q=15/14$, as predicted by Eqs.~\eqref{eq:su2} and \eqref{eq:transmission-ratio}.

For $\rho=1$, $\kk=9/8$ exceeds $k_L=1$.
The reflection entry of the overlap matrix \eqref{eq:current-overlap} for incidence from the left is $S_{LL}=1-\kk/k_L=-1/8$, so the reflected current component has the opposite sign to the incident one \cite{KM2015}. Energy does not behave in this way. The transmitted and reflected energy fractions are $\cc/c_L=27/32$ and $1-\cc/c_L=5/32$, both positive. Thus $\kk$ is not bounded by the smaller bulk level, while $\cc$ is bounded by the smaller central charge. The values $\kk=9/8$ and $\cc=27/32$ are consistent with the bounds in Eqs.~\eqref{eq:k-strong-bound} and~\eqref{eq:c-strong-bound}, which give $\kk\le5/4$ and $\cc\le15/16$.

\subsection{\texorpdfstring{$SU(3)$}{SU(3)} examples}\label{sec:su3}

The coset construction also gives transmitting $SU(3)$ interfaces at unequal and equal levels. For an interface between $SU(3)_1$ and $SU(3)_2$, choose $\rho$ to be the fundamental representation of the diagonal $SU(3)_3$ algebra. Equation~\eqref{eq:generalomega} then gives $\omega_\rho=0$. Appendix~\ref{app:su3} gives the character calculation. Equations~\eqref{eq:komega} and \eqref{eq:comega} yield
\begin{equation}
\begin{aligned}
\kk&=\frac23,\qquad \cc=\frac{32}{27},\\
\mathcal T_Q&=\frac49,\qquad \mathcal T_E=\frac{160}{351}.
\end{aligned}
\label{eq:su3example}
\end{equation}
Here $\cc/\kk=16/9$ and $\mathcal T_E/\mathcal T_Q=40/39$, as in Eqs.~\eqref{eq:main} and \eqref{eq:transmission-ratio}.

For an interface between two $SU(3)_1$ models, choose $\rho$ to be the fundamental representation of the diagonal $SU(3)_2$ algebra. The same character calculation now gives $\omega_\rho=-(3-\sqrt5)/2$, hence
\begin{equation}
\kk=\frac{5-\sqrt5}{4},\qquad \cc=\frac{5-\sqrt5}{2}.
\label{eq:su3equal}
\end{equation}
Both bulk central charges equal $2$, giving $\mathcal T_E=\mathcal T_Q=(5-\sqrt5)/4$.

\section{Constraints on thermal and charge conductances}\label{sec:conductance}

For WZW bulk theories, diagonal affine symmetry fixes the ratio of thermal to charge conductance for $G_Q>0$ even when different interfaces have different transmission coefficients. With bulk coset sectors, their central charges instead bound the thermal conductance at fixed temperature and charge conductance. Current correlations determine charge conductance in boundary conformal field theory \cite{ChamonOA,OCA2006,Rahmani}, and thermal and charge transport through impurities and free-boson wire junctions have been studied \cite{KaneFisher,GT}.

Let $X$ be a unit-norm generator of the preserved algebra and $Q$ its charge, with $J_i^X=X_aJ_i^a$ in the basis of Eq.~\eqref{eq:ope}. Write $G_Q$ and $\kappa$ for the charge and thermal conductances in linear response at zero chemical potential, and set $\hbar=k_B=1$.

In the unfolded picture, the charge and energy currents normal to the interface on side $i$ are
\begin{equation*}
j_Q=\frac{J_i^X-\bar J_i^X}{2\pi},\qquad j_E=\frac{T_i-\bar T_i}{2\pi}.
\end{equation*}
For the twisted gluing of Eq.~\eqref{eq:twisted-gluing}, the conserved charge current on side $i$ is instead
\begin{equation*}
j_Q=\frac{[R_i(J_i-\Omega_i\bar J_i)]^X}{2\pi},
\end{equation*}
with $\kk$ defined in the basis of the preserved affine algebra. Current and conformal gluing, together with Eq.~\eqref{eq:coeff} in that basis, give
\begin{equation}
\begin{aligned}
\langle j_Q(\tau)j_Q(0)\rangle_c&=\frac{\kk}{2\pi^2\tau^2},\\
\langle j_E(\tau)j_E(0)\rangle_c&=\frac{\cc}{4\pi^2\tau^4}.
\end{aligned}
\label{eq:normalcorr}
\end{equation}
The subscript $c$ denotes the connected correlator. Mapping to the thermal cylinder and applying the equilibrium fluctuation--dissipation theorem gives
\begin{equation}
G_Q=\frac{\kk}{2\pi},\qquad \frac{\kappa}{T}=\frac{\pi}{6}\cc.
\label{eq:kubo}
\end{equation}
Appendix~\ref{app:kubo} gives the noise normalization.
Restoring units, with $q$ the physical charge per unit of $Q$,
\begin{equation}
G_{qQ}=\kk\,\frac{q^2}{h},\qquad
\kappa=\cc\,\frac{\pi^2k_B^2T}{3h}.
\label{eq:units}
\end{equation}
Thus $\kk$ and $\cc$ give the charge and thermal conductances in units of their respective conductance quanta \cite{Pendry,RegoKirczenow,Schwab,Jezouin}.

At zero chemical potential the interface has no thermoelectric response at the conformal point. With respect to the total Virasoro algebra preserved by Eq.~\eqref{eq:conformal-gluing}, $J_i^X$ and $T_i$ are quasiprimary fields of weights one and two, so global conformal symmetry along the interface sets to zero every two-point function between them at separated points, of either chirality and on either side. The Schwarzian term of the map to the cylinder only shifts one-point functions, which cancel in $j_E$. Hence $\langle j_E(\tau)j_Q(0)\rangle_c=0$ at any temperature and for any interface preserving $Q$, including the $U(1)$ interfaces of Eq.~\eqref{eq:u1-window}, and $\kappa$ is also the open-circuit thermal conductance. Thermopower can arise from irrelevant operators that break particle--hole symmetry, as for a Luttinger liquid \cite{KaneFisher}.

Combining Eqs.~\eqref{eq:main} and \eqref{eq:kubo} gives, for $G_Q>0$,
\begin{equation}
\frac{\kappa}{T G_Q}=\frac{\pi^2}{3}\frac{2\dim G}{k_L+k_R+2h^\vee}.
\label{eq:lorenz}
\end{equation}
The coefficient $\kk$ cancels, so different interfaces can have different conductances but the same ratio.
Equation~\eqref{eq:lorenz} is a generalized Wiedemann--Franz law.
For a homogeneous $G_{k_i}$ WZW model, the corresponding ratio is $\pi^2c_i/(3k_i)$, the analog of the ideal spinless Luttinger-liquid value $\pi^2/(3g)$ for Luttinger parameter $g$ \cite{KaneFisher}. The ratio in Eq.~\eqref{eq:lorenz} is the harmonic mean of the two bulk ratios,
\begin{equation}
\frac{\kappa}{TG_Q}=\frac{2\pi^2}{3(k_L/c_L+k_R/c_R)}.
\label{eq:harmonic}
\end{equation}
At equal levels this equals the homogeneous ratio.
The conductances of the two-wire free-boson junction \cite{GT} give the same harmonic mean (see below). Here it follows for any transmitting interface between WZW models preserving the diagonal affine currents. For $SU(2)_1$ and $SU(2)_3$, the bulk ratios are $\pi^2/3$ and $\pi^2/5$. The two transmitting interfaces in Table~\ref{tab:su2} differ in both $G_Q$ and $\kappa$ by a factor of $3/2$, yet both have
\begin{equation}
\frac{\kappa}{T G_Q}=\frac{\pi^2}{4}
\label{eq:su2-lorenz}
\end{equation}
for a unit-norm generator.

The charge normalization matters for a measured conductance. In the standard $SU(2)$ convention, the current for $S^3$ is $J_i^3/\sqrt{2}$ in the normalization of Eq.~\eqref{eq:ope}, so the coefficient of the double pole in its OPE with itself is $k_i/2$. For the physical spin component $S^3$, this gives $G_{S^3}=\kk/(4\pi)$.
For $SU(2)$ interfaces Eq.~\eqref{eq:lorenz} becomes
\begin{equation}
\frac{\kappa}{T G_{S^3}}=\frac{4\pi^2}{k_L+k_R+4},
\label{eq:su2-spin-lorenz}
\end{equation}
which equals $\pi^2/2$ for both transmitting rows of Table~\ref{tab:su2}.
An electrical Lorenz ratio likewise depends on the normalization of the electric charge.

For a single preserved diagonal $U(1)$ current with levels $k_L$ and $k_R$, and free-boson bulk theories, $T_i={:J_iJ_i:}/(2k_i)$ is the full stress tensor and the derivation of Eq.~\eqref{eq:main} applies with $h^\vee=0$ and $\dim G=1$, giving
\begin{equation}
\cc=\frac{2\kk}{k_L+k_R}.
\label{eq:u1-relation}
\end{equation}
A free Dirac fermion of unit charge with transmission probability $\mathcal T$ has $\kk=\cc=\mathcal T$, consistent with Eq.~\eqref{eq:u1-relation}. For spinless Luttinger liquids, $k_i=g_i$ \cite{KaneFisher}. The healed junction of two semi-infinite Luttinger liquids has intrinsic electrical conductance \cite{Hou}
\begin{equation*}
G_{eQ}=g_{\rm eff}\frac{e^2}{h},\qquad g_{\rm eff}^{-1}=\frac{g_L^{-1}+g_R^{-1}}{2}.
\end{equation*}
It saturates the bound \eqref{eq:unitarity-bounds}, $\kk\le2k_Lk_R/(k_L+k_R)=g_{\rm eff}$. Equations~\eqref{eq:units} and \eqref{eq:u1-relation} then give
\begin{equation}
\begin{aligned}
\cc&=\frac{4g_Lg_R}{(g_L+g_R)^2}=1-\gamma^2,\\
\frac{\kappa}{TG_{eQ}}&=\frac{\pi^2 k_B^2}{3e^2}\frac{2}{g_L+g_R},
\end{aligned}
\label{eq:LL}
\end{equation}
with $\gamma=(g_L-g_R)/(g_L+g_R)$ the plasmon reflection amplitude for incidence from the left \cite{Safi}, in agreement with the thermal conductance of the free-boson junction \cite{GT}. Contact with external leads can change the measured electrical conductance \cite{Safi,MaslovStone,Ponomarenko}. Since $\cc=\kk^2/(k_Lk_R)$ here, this junction also saturates Eq.~\eqref{eq:quadratic-lower}. The unitarity and ANEC-type bounds meet at this $U(1)$ interface.

With bulk coset stress tensors, Eq.~\eqref{eq:coset-window} and Eq.~\eqref{eq:kubo} give a range for the thermal conductance at fixed temperature and charge conductance. For $c_L^\perp\ge c_R^\perp$,
\begin{align}
0&\le\frac{\kappa}{T}-\frac{2\pi^2\dim G}{3(k_L+k_R+2h^\vee)}G_Q\nonumber\\
&\le\frac{\pi}{6}\Bigl[c_R^\perp+\mu_R^2\min\bigl(c_L^\perp-c_R^\perp,\,c_0\bigr)\Bigr],
\label{eq:conductance-range}
\end{align}
with $\mu_R$ and $c_0$ defined in Eqs.~\eqref{eq:mu-def} and \eqref{eq:T0-def}, and with $L\leftrightarrow R$ otherwise.
Thus bulk coset sectors can carry additional heat at fixed charge conductance, but the increase in $\kappa/T$ is bounded by their central charges. The bound also applies when charge is completely reflected. For example, an $SU(2)_2$ interface preserving the Cartan current has $\kappa/T\le\pi/12$ when $G_Q=0$ by Eq.~\eqref{eq:u1-window}. The Majorana interface of Eq.~\eqref{eq:majorana-window} saturates this limit by transmitting $\chi_3$.

The coefficient $\kk$ also fixes the leading equilibrium charge fluctuations. Integrating the symmetrized real-time correlator obtained from Eq.~\eqref{eq:normalcorr} over a time window of length $t$ gives the variance of the charge $\Delta Q(t)$ transferred through the interface at zero temperature. For $t\gg\epsilon$,
\begin{equation}
\operatorname{Var}[\Delta Q(t)]=\frac{\kk}{\pi^2}\ln\frac{t}{\epsilon}+O(1),
\label{eq:FCS}
\end{equation}
with $\epsilon$ a short-time cutoff. For $T\epsilon\ll1$, the finite-temperature variance follows from Eq.~\eqref{eq:FCS} with the substitution
\begin{equation*}
\ln\frac{t}{\epsilon}\longrightarrow\ln\!\left[\frac{\sinh(\pi Tt)}{\pi T\epsilon}\right].
\end{equation*}
At long times $t\gg T^{-1}$, the variance crosses over to the Johnson--Nyquist growth $2TG_Q\,t$. For a free fermion, $\kk$ is the transmission probability and Eq.~\eqref{eq:FCS} reproduces the logarithmic growth of the second cumulant of the transferred charge at a quantum point contact \cite{LevitovLesovik,LLL,Schoenhammer,Klich}.

\section{Transmission and boundary-condition-changing operators}\label{sec:twists}

The coefficient $\kk$ also appears in the boundary spectrum. Rotate the right bulk theory by a symmetry transformation on a finite segment of the interface. In the unfolded picture, this places a segment of a symmetry line of the right bulk theory along the interface. In the folded picture, the boundary condition changes at the two endpoints of the segment, and a boundary-condition-changing (BCC) operator is placed at each endpoint. We show below that its scaling dimension $h_{\rm BCC}(\theta)$ at rotation angle $\theta$ satisfies $h_{\rm BCC}''(0)=\kk/(2\pi^2)$. For WZW bulk theories, Eq.~\eqref{eq:main} then gives $\cc$ as well. Assume that the boundary spectrum is discrete and has a unique identity operator. On a strip of width $\ell$ with the boundary condition $|B\rangle$ on one side and the rotated one on the other, the ground-state energy for sufficiently small $\theta$ is shifted by $\pi h_{\rm BCC}(\theta)/\ell$ relative to $\theta=0$, in units with velocity one. Thus $\kk$, and $\cc$ for WZW bulk theories, can be obtained from the finite-size spectrum of a lattice model.

Start with the untwisted gluing condition \eqref{eq:gluing} and rotate both chiralities of the right bulk theory by $e^{i\theta X}$, where $X$ is a unit-norm generator, only on a segment $(s,t)$ of the interface. This relative rotation leaves $\cc$ unchanged. Denote the scaling dimension by $h_{\rm BCC}(\theta)$. We follow the branch with $h_{\rm BCC}(0)=0$.

An infinitesimal rotation inserts $\theta\int_s^t j_Q(\tau)\,d\tau$, with $j_Q=(J_R^X-\bar J_R^X)/(2\pi)$. Its second-order contribution to the logarithm of the BCC two-point function is a double integral of the connected current correlator in Eq.~\eqref{eq:normalcorr}. The two-point function of the boundary-condition-changing operators scales as $(t-s)^{-2h_{\rm BCC}(\theta)}$ at separations large compared with a short-distance cutoff. Applying $-\partial_s\partial_t$ to the logarithm removes endpoint normalization and terms proportional to the segment length, giving
\begin{equation}
\frac{2h_{\rm BCC}(\theta)}{(t-s)^2}=\theta^2\langle j_Q(s)j_Q(t)\rangle_c+o(\theta^2).
\label{eq:bcc-endpoint-derivatives}
\end{equation}
Using Eq.~\eqref{eq:normalcorr} gives
\begin{align}
h_{\rm BCC}(\theta)&=\frac{\kk}{4\pi^2}\theta^2+o(\theta^2)\nonumber\\
&=\frac{k_L+k_R+2h^\vee}{8\pi^2\dim G}\cc\theta^2+o(\theta^2).
\label{eq:bcc}
\end{align}
The second equality holds for WZW bulk theories. By Eq.~\eqref{eq:kubo}, $h_{\rm BCC}''(0)=G_Q/\pi$. The second derivative of the scaling dimension at zero rotation angle determines the charge conductance in linear response. For WZW bulk theories, it also determines $\cc$ and $\kappa/T$ through Eqs.~\eqref{eq:main} and \eqref{eq:kubo}. With bulk coset stress tensors, the relation to charge conductance still holds, and Eq.~\eqref{eq:conductance-range} bounds $\kappa/T$ at fixed $h_{\rm BCC}''(0)$.

For automorphism-twisted gluing, the rotation is generated by the right-hand part of the preserved charge, using the right-hand current $j_Q$ of Sec.~\ref{sec:conductance} in the basis of the preserved affine algebra. The same derivation then applies. Rotations of a single WZW boundary were studied in Ref.~\cite{ChoiEtAl}. Here the rotation acts on one factor in the folded picture. The same scaling dimension governs the long-time power law of the overlap with the initial ground state when the relative rotation is switched on and off \cite{AffleckBCC}.

For the two transmitting rows of Table~\ref{tab:su2}, the dimensions are
\begin{equation}
h_{\rm BCC}(\theta)=
\begin{cases}
3\theta^2/(16\pi^2)+o(\theta^2),&\rho=\frac12,\\
9\theta^2/(32\pi^2)+o(\theta^2),&\rho=1.
\end{cases}
\end{equation}
Two limits check Eq.~\eqref{eq:bcc}. For a factorizing interface with diagonal current gluing, each side preserves the symmetry, the rotation acts trivially on $|B\rangle$, and $h_{\rm BCC}=0$, in agreement with $\kk=0$. For the identity interface between identical theories, $\kk=k$ and Eq.~\eqref{eq:bcc} gives $h_{\rm BCC}=k\theta^2/(4\pi^2)+o(\theta^2)$. In this case the rotation inserts a segment of the symmetry line $e^{i\theta Q}$ in a single bulk theory \cite{GKSW,CLSWY}. Its endpoints are twist fields with $h=\bar h=k\theta^2/(8\pi^2)$ to quadratic order, and $h+\bar h$ agrees with $h_{\rm BCC}$. At finite angle, the scaling dimension depends on the boundary spectrum in the folded picture beyond $\kk$.

\section{Discussion}

An interface that carries heat while reflecting a conserved charge would allow energy and charge transport to be controlled separately. Kimura and Murata proposed reducing $SU(2)$ symmetry to $U(1)$ to study spin transport \cite{KM2014}, and emphasized the need to identify the microscopic interactions that realize particular interface conditions \cite{KM2015}. Our bounds quantify the separation this symmetry reduction permits. For $SU(2)_2$ bulk theories at zero chemical potential, blocking the preserved spin current limits the thermal conductance to one third of its fully transmitting value. A concrete next step is to vary local spin-chain couplings that conserve a chosen spin component and determine which interactions approach this limit \cite{AffleckHaldane,ALKondo1991a,ALKondo1991b}.

The bounds also identify when additional bulk degrees of freedom are required. At unequal current levels, Eq.~\eqref{eq:coset-cost} bounds the coset central charge required on the higher-level side for complete energy transmission from the lower-level WZW theory. This is the part of the bulk central charge not accounted for by the preserved current algebra. If it is too small, tuning the interface interaction cannot restore complete transmission while preserving the diagonal currents. Thus the bound can rule out a proposed bulk theory before a detailed search for interface interactions. The next step is to identify bulk theories and local interface interactions that achieve complete transmission with the smallest possible coset central charge.

\begin{acknowledgments}
YK is supported by the INAMORI Frontier Program at Kyushu University, JSPS KAKENHI Grant Number 23K20046, MEXT Supporting Pioneering Research through AI for 1,000 Discovery Challenges (SPReAD) Grant Number JPMXP1726262823, and JST PRESTO Grant Number JPMJPR26A6.
TO is supported by JST SPRING, Grant Number JPMJSP2110.
We used \mbox{ChatGPT} (OpenAI, GPT-6.2 Astra) and Claude (Anthropic, Opus 5.5 and Fable 5.2) to assist with drafting and checking the manuscript under our direction. The authors are responsible for all content and conclusions.
\end{acknowledgments}

\appendix
\section{Current states and the energy bound}\label{app:bound}

For WZW bulk theories, Eq.~\eqref{eq:energy-current-bound} follows from the current OPE and positivity of outgoing averaged null energy. In the folded picture, the incoming current $J_R^a$ on the right corresponds to $\bar J_R^a$ in the unfolded picture. First, the current $J_i^a$ has conformal weight one. With no sum over $a$, the only $G$-singlet spin-two term in the current OPE is
\begin{equation}
J_i^a(z)J_i^a(0)\supset\frac{2k_i}{c_i}T_i(0).
\label{eq:current-singlet}
\end{equation}
The stress-tensor Ward identity fixes this coefficient. Diagonal $G$ symmetry removes nonsinglet spin-two contributions to the stress-tensor correlator across the interface. The ratio of the relevant three-point functions is $\cc/c_i$. Using the interface scattering construction of Ref.~\cite{MPR}, we identify this ratio with the transmitted energy fraction of an incoming current wave packet.

Second, choose incoming and outgoing wave packets of unit norm created by a current, with the same energy profile and Lie-algebra index, on opposite sides. Take the limit of vanishing energy width at fixed $E>0$. Their overlap is $\kk/\sqrt{k_Lk_R}$. Let $\mathcal E_j$ be the averaged null energy operator measuring outgoing energy on side $j$. The outgoing current state carries energy $E$ entirely on side $j$, so it is an eigenstate of $\mathcal E_j$ in this limit. Positivity of $\mathcal E_j$ on the orthogonal complement gives
\begin{equation}
\frac{\kk^2}{k_Lk_R}\le\frac{\langle\mathrm{in},i,a;E|\mathcal E_j|\mathrm{in},i,a;E\rangle}{E}=\frac{\cc}{c_i},\qquad j\ne i.
\label{eq:energy-projection}
\end{equation}
Applying Eq.~\eqref{eq:energy-projection} from both sides yields Eq.~\eqref{eq:energy-current-bound}.

\section{Equilibrium response and charge normalization}\label{app:kubo}

We work at zero chemical potential in the basis of the preserved algebra. In the unfolded picture, for local currents on either side of the interface and a unit-norm generator $X$, the correlator of opposite chiralities on side $i$ has coefficient $k_i-\kk$ at $x=0$,
\begin{equation}
\begin{aligned}
\langle J_i^X(\tau)J_i^X(0)\rangle&=\frac{k_i}{\tau^2},\\
\langle J_i^X(\tau)\bar J_i^X(0)\rangle&=\frac{k_i-\kk}{\tau^2}.
\end{aligned}
\label{eq:reflocal}
\end{equation}
The same coefficient occurs with $J_i$ and $\bar J_i$ exchanged. Thus $j_Q=(J_i^X-\bar J_i^X)/(2\pi)$ has the first correlator in Eq.~\eqref{eq:normalcorr}. For a totally reflecting interface, $\kk=0$ and $j_Q$ vanishes on the interface. The opposite-chirality stress-tensor correlator has coefficient $(c_i-\cc)/2$, giving the second. The minus sign in the boundary-state overlap in Eq.~\eqref{eq:coeff} does not enter the local expression for $j_Q$.

At temperature $T$ the conformal map from the plane to a cylinder of circumference $1/T$ replaces the factors in the connected plane correlators by
\begin{equation}
\frac{1}{\tau^{2n}}\longrightarrow\left(\frac{\pi T}{\sin(\pi T\tau)}\right)^{2n},\qquad n=1,2.
\label{eq:cylinder}
\end{equation}
The symmetrized zero-frequency current noise in equilibrium is
\begin{equation}
\mathcal S_A(0)=\frac12\int_{-\infty}^{\infty}dt\,\langle\{\delta j_A(t),\delta j_A(0)\}\rangle,\qquad A=Q,E,
\label{eq:noise-def}
\end{equation}
with $\delta j_A=j_A-\langle j_A\rangle$.
Set $\tau=\epsilon+it$. With $a=\pi T$, the integrals
\begin{equation}
\begin{aligned}
\int_{-\infty}^{\infty}\frac{dt}{\sinh^2[a(t-i\epsilon)]}&=-\frac{2}{a},\\
\int_{-\infty}^{\infty}\frac{dt}{\sinh^4[a(t-i\epsilon)]}&=\frac{4}{3a}
\end{aligned}
\label{eq:noise-integrals}
\end{equation}
are independent of $\epsilon$ for $0<\epsilon<1/T$, and both operator orderings give the same result. Including the factors $(-1)^n$ from the analytic continuation yields
\begin{equation}
\mathcal S_Q(0)=\frac{\kk T}{\pi},\qquad \mathcal S_E(0)=\frac{\pi\cc T^3}{3}.
\label{eq:noise}
\end{equation}
The fluctuation--dissipation relations $\mathcal S_Q(0)=2TG_Q$ and $\mathcal S_E(0)=2T^2\kappa$ reproduce Eq.~\eqref{eq:kubo}.
At $T=0$, the same continuation, integrated over a time window $t\gg\epsilon$, gives Eq.~\eqref{eq:FCS}.

For a generator $X$ of arbitrary norm, the corresponding current correlator coefficient is $\bigl(\sum_a X_a^2\bigr)\kk$. Accordingly, multiply the right-hand sides of the charge-conductance formula in Eq.~\eqref{eq:kubo}, of Eq.~\eqref{eq:FCS}, and the quadratic term of Eq.~\eqref{eq:bcc} by $\sum_a X_a^2$. The coefficient $\cc$ has no analogous generator normalization.

For an identity interface, rotating the right bulk factor changes both chiral pairings in the folded picture. The quadratic coefficient of the scaling dimension in Eq.~\eqref{eq:bcc} is consequently twice that for a rotation of a single WZW boundary generated by the holomorphic current, as studied in Ref.~\cite{ChoiEtAl}.

\section{Modular \texorpdfstring{$S$}{S}-matrix check for the \texorpdfstring{$SU(3)$}{SU(3)} examples}\label{app:su3}

The state $\mathcal K_{-1}^a|0\rangle$ transforms in the adjoint representation of the diagonal $SU(3)_K$ algebra. For the interfaces of Sec.~\ref{sec:su3}, its boundary-state coefficient gives Eq.~\eqref{eq:generalomega} \cite{QS}. We evaluate this ratio for the fundamental representation $\rho=(1,0)$ of the diagonal algebra at $K=3$ and $K=2$.

The ratio of modular $S$-matrix elements at level $K$ equals a finite-dimensional character,
\begin{equation}
\frac{S^{(K)}_{\rho,{\rm adj}}}{S^{(K)}_{\rho0}}=\chi_{\rm adj}\!\left(e^{-2\pi i(\rho+\varrho)/(K+3)}\right),
\label{eq:Scharacter}
\end{equation}
where $\varrho$ is the Weyl vector. Since $\mathbf 3\otimes\overline{\mathbf 3}=\mathbf 1\oplus\mathbf 8$, the adjoint character is $\chi_{\rm adj}=|\chi_{\mathbf 3}|^2-1$ on the maximal torus of $SU(3)$. For the fundamental $\rho=(1,0)$ at $K=3$, the three phase differences between eigenvalues in the fundamental representation are $\pi/3$, $2\pi/3$, and $\pi$. Consequently,
\begin{equation}
|\chi_{\mathbf 3}|^2=3+2\left[\cos\frac\pi3+\cos\frac{2\pi}3+\cos\pi\right]=1,
\end{equation}
so $\chi_{\rm adj}=0$. At $\rho=0$, the phase differences are $\pi/3$, $\pi/3$, and $2\pi/3$, yielding $\chi_{\rm adj}=3$. Thus $\omega_{(1,0)}=0$, giving Eq.~\eqref{eq:su3example}.

For $K=2$ and $\rho=(1,0)$, the phase differences are $2\pi/5$, $4\pi/5$, and $6\pi/5$. They give $\chi_{\rm adj}=-(\sqrt5-1)/2$. For the vacuum representation $\rho=0$, the phase differences are $2\pi/5$, $2\pi/5$, and $4\pi/5$, giving $\chi_{\rm adj}=(1+\sqrt5)/2$. Their ratio is $\omega_{(1,0)}=-(3-\sqrt5)/2$, giving Eq.~\eqref{eq:su3equal} at $k_L=k_R=1$.


\begin{thebibliography}{99}
\bibitem{FFRS2004} J. Fr\"ohlich, J. Fuchs, I. Runkel, and C. Schweigert, ``Kramers-Wannier duality from conformal defects,'' Phys. Rev. Lett. \textbf{93}, 070601 (2004), \href{https://arxiv.org/abs/cond-mat/0404051}{cond-mat/0404051}.
\bibitem{QRW} T. Quella, I. Runkel, and G. M. T. Watts, ``Reflection and transmission for conformal defects,'' J. High Energy Phys. \textbf{04} (2007) 095, \href{https://arxiv.org/abs/hep-th/0611296}{hep-th/0611296}.
\bibitem{SakaiSatoh} K. Sakai and Y. Satoh, ``Entanglement through conformal interfaces,'' J. High Energy Phys. \textbf{12} (2008) 001, \href{https://arxiv.org/abs/0809.4548}{arXiv:0809.4548}.
\bibitem{KRW2009} M. Kormos, I. Runkel, and G. M. T. Watts, ``Defect flows in minimal models,'' J. High Energy Phys. \textbf{11} (2009) 057, \href{https://arxiv.org/abs/0907.1497}{arXiv:0907.1497}.
\bibitem{Gaiotto2012} D. Gaiotto, ``Domain walls for two-dimensional renormalization group flows,'' J. High Energy Phys. \textbf{12} (2012) 103, \href{https://arxiv.org/abs/1201.0767}{arXiv:1201.0767}.
\bibitem{KM2014} T. Kimura and M. Murata, ``Current reflection and transmission at conformal defects: Applying BCFT to transport process,'' Nucl. Phys. B \textbf{885}, 266 (2014), \href{https://arxiv.org/abs/1402.6705}{arXiv:1402.6705}.
\bibitem{BrehmBrunner} E. M. Brehm and I. Brunner, ``Entanglement entropy through conformal interfaces in the 2D Ising model,'' J. High Energy Phys. \textbf{09} (2015) 080, \href{https://arxiv.org/abs/1505.02647}{arXiv:1505.02647}.
\bibitem{BrunnerSchmidt2016} I. Brunner and C. Schmidt-Colinet, ``Reflection and transmission of conformal perturbation defects,'' J. Phys. A \textbf{49}, 195401 (2016), \href{https://arxiv.org/abs/1508.04350}{arXiv:1508.04350}.
\bibitem{MakabeWatts} I. Makabe and G. M. T. Watts, ``The reflection coefficient for minimal model conformal defects from perturbation theory,'' J. High Energy Phys. \textbf{06} (2018) 143, \href{https://arxiv.org/abs/1712.07234}{arXiv:1712.07234}.
\bibitem{Bachas2020} C. Bachas, S. Chapman, D. Ge, and G. Policastro, ``Energy reflection and transmission at 2D holographic interfaces,'' Phys. Rev. Lett. \textbf{125}, 231602 (2020), \href{https://arxiv.org/abs/2006.11333}{arXiv:2006.11333}.
\bibitem{Bachas2023} C. Bachas, S. Baiguera, S. Chapman, G. Policastro, and T. Schwartzman, ``Energy transport for thick holographic branes,'' Phys. Rev. Lett. \textbf{131}, 021601 (2023), \href{https://arxiv.org/abs/2212.14058}{arXiv:2212.14058}.
\bibitem{KKOSW2023} A. Karch, Y. Kusuki, H. Ooguri, H.-Y. Sun, and M. Wang, ``Universality of effective central charge in interface CFTs,'' J. High Energy Phys. \textbf{11} (2023) 126, \href{https://arxiv.org/abs/2308.05436}{arXiv:2308.05436}.
\bibitem{KKOSW2024} A. Karch, Y. Kusuki, H. Ooguri, H.-Y. Sun, and M. Wang, ``Universal bound on effective central charge and its saturation,'' Phys. Rev. Lett. \textbf{133}, 091604 (2024), \href{https://arxiv.org/abs/2404.01515}{arXiv:2404.01515}.
\bibitem{Furuta2026} Y. Furuta, Y. Kusuki, and T. Onagi, ``Transmission coefficients from phantom currents,'' Phys. Rev. D \textbf{113}, 045008 (2026), \href{https://arxiv.org/abs/2511.00356}{arXiv:2511.00356}.
\bibitem{BaigKarch2022} S. A. Baig and A. Karch, ``Double brane holographic model dual to 2d ICFTs,'' J. High Energy Phys. \textbf{10} (2022) 022, \href{https://arxiv.org/abs/2206.01752}{arXiv:2206.01752}.
\bibitem{BaigShashi2023} S. A. Baig and S. Shashi, ``Transport across interfaces in symmetric orbifolds,'' J. High Energy Phys. \textbf{10} (2023) 168, \href{https://arxiv.org/abs/2301.13198}{arXiv:2301.13198}.
\bibitem{BaigKarchWang2024} S. A. Baig, A. Karch, and M. Wang, ``Transmission coefficient of super-Janus solution,'' J. High Energy Phys. \textbf{10} (2024) 235, \href{https://arxiv.org/abs/2408.00059}{arXiv:2408.00059}.
\bibitem{GutperleHultgreenMena2025} M. Gutperle and C. Hultgreen-Mena, ``Janus and RG-interfaces in minimal 3d gauged supergravity,'' Int. J. Mod. Phys. A \textbf{40}, 2550113 (2025), \href{https://arxiv.org/abs/2412.16749}{arXiv:2412.16749}.
\bibitem{BaradTangWen2025} R. Barad, Q. Tang, and X. Wen, ``Dissipation meets conformal interface in open quantum systems: How the relaxation rate is suppressed,'' Phys. Rev. B \textbf{112}, 235143 (2025), \href{https://arxiv.org/abs/2505.04715}{arXiv:2505.04715}.
\bibitem{AntinucciEtAl2025} A. Antinucci, C. Copetti, G. Galati, and G. Rizi, ``Defect conformal manifolds from phantom noninvertible symmetries,'' Phys. Rev. Lett. \textbf{135}, 211602 (2025), \href{https://arxiv.org/abs/2505.09668}{arXiv:2505.09668}.
\bibitem{LiuWang2025} Y. Liu and C.-Y. Wang, ``Energy transport in holographic junctions,'' J. High Energy Phys. \textbf{10} (2025) 205, \href{https://arxiv.org/abs/2506.19553}{arXiv:2506.19553}.
\bibitem{AbdallaChandraWang2026} A. I. Abdalla, J. Chandra, and Y. Wang, ``Line defects in Liouville conformal field theory: Localized cosmological constants and decohered hyperbolic geometries,'' J. High Energy Phys. \textbf{10} (2026) 165, \href{https://arxiv.org/abs/2603.02166}{arXiv:2603.02166}.
\bibitem{ZhangEtAl2026} J.-R. Zhang, J.-H. Jin, T.-K. Chen, and J. Chen, ``Defect conformal manifolds along RG domain walls between $\mathbb{Z}_N$-parafermions and minimal models,'' J. High Energy Phys. \textbf{09} (2026) 123, \href{https://arxiv.org/abs/2605.24978}{arXiv:2605.24978}.
\bibitem{FurutaEtAlComplex2026} Y. Furuta, W. Harada, Y. Kusuki, and Y. Tang, ``Complex conformal manifolds,'' \href{https://arxiv.org/abs/2606.30720}{arXiv:2606.30720}.
\bibitem{AravEtAl2026} I. Arav, T. Bertrand, S. Chapman, G. Policastro, and S. Waeber, ``Energy transmission across holographic conformal interfaces in general dimensions,'' \href{https://arxiv.org/abs/2609.38318}{arXiv:2609.38318}.
\bibitem{ChenLogarithmic2026} J. Chen, ``When conformal interfaces become logarithmic,'' \href{https://arxiv.org/abs/2610.23509}{arXiv:2610.23509}.
\bibitem{ChenRGWalls2026} J. Chen, ``Renormalization group domain walls from phantom symmetries,'' \href{https://arxiv.org/abs/2610.23529}{arXiv:2610.23529}.
\bibitem{AntunesChenRong2026} A. Antunes, J. Chen, and J. Rong, ``Rational conformal interfaces between the Ising and tricritical Ising models from a tetrality web,'' \href{https://arxiv.org/abs/2611.18831}{arXiv:2611.18831}.
\bibitem{Zheng2026} B. Zheng, ``Classification of conformal Ising/tricritical Ising interfaces preserving $\mathcal{W}_3^{\mathbb{Z}_2}$ symmetry,'' \href{https://arxiv.org/abs/2611.19129}{arXiv:2611.19129}.
\bibitem{LiuWangZeng2025} Y. Liu, C.-Y. Wang, and Y.-J. Zeng, ``Energy transport in holographic non-conformal interfaces,'' J. High Energy Phys. \textbf{09} (2025) 143, \href{https://arxiv.org/abs/2503.20399}{arXiv:2503.20399}.
\bibitem{BanerjeePolicastro2025} A. Banerjee and G. Policastro, ``Energy reflection and transmission of interfaces in $T\bar{T}$-deformed CFT,'' \href{https://arxiv.org/abs/2512.03167}{arXiv:2512.03167}.
\bibitem{Billo2016} M. Bill\`o, V. Gon\c{c}alves, E. Lauria, and M. Meineri, ``Defects in conformal field theory,'' J. High Energy Phys. \textbf{04} (2016) 091, \href{https://arxiv.org/abs/1601.02883}{arXiv:1601.02883}.
\bibitem{Andrei2020} N. Andrei, A. Bissi, M. Buican, J. Cardy, P. Dorey, N. Drukker, J. Erdmenger, D. Friedan, D. Fursaev, A. Konechny, C. Kristjansen, I. Makabe, Y. Nakayama, A. O'Bannon, R. Parini, B. Robinson, S. Ryu, C. Schmidt-Colinet, V. Schomerus, C. Schweigert, and G. M. T. Watts, ``Boundary and defect CFT: open problems and applications,'' J. Phys. A \textbf{53}, 453002 (2020), \href{https://arxiv.org/abs/1810.05697}{arXiv:1810.05697}.
\bibitem{MPR} M. Meineri, J. Penedones, and A. Rousset, ``Colliders and conformal interfaces,'' J. High Energy Phys. \textbf{02} (2020) 138, \href{https://arxiv.org/abs/1904.10974}{arXiv:1904.10974}.
\bibitem{BPZ} A. A. Belavin, A. M. Polyakov, and A. B. Zamolodchikov, ``Infinite conformal symmetry in two-dimensional quantum field theory,'' Nucl. Phys. B \textbf{241}, 333 (1984).
\bibitem{Witten1984} E. Witten, ``Nonabelian bosonization in two dimensions,'' Commun. Math. Phys. \textbf{92}, 455 (1984).
\bibitem{KZ1984} V. G. Knizhnik and A. B. Zamolodchikov, ``Current algebra and Wess--Zumino model in two dimensions,'' Nucl. Phys. B \textbf{247}, 83 (1984).
\bibitem{DiFrancesco} P. Di Francesco, P. Mathieu, and D. S\'en\'echal, \emph{Conformal Field Theory} (Springer, New York, 1997).
\bibitem{KM2015} T. Kimura and M. Murata, ``Transport process in multi-junctions of quantum systems,'' J. High Energy Phys. \textbf{07} (2015) 072, \href{https://arxiv.org/abs/1505.05275}{arXiv:1505.05275}.
\bibitem{KaneFisher} C. L. Kane and M. P. A. Fisher, ``Thermal transport in a Luttinger liquid,'' Phys. Rev. Lett. \textbf{76}, 3192 (1996), \href{https://arxiv.org/abs/cond-mat/9509068}{cond-mat/9509068}.
\bibitem{Fazio} R. Fazio, F. W. J. Hekking, and D. E. Khmelnitskii, ``Anomalous thermal transport in quantum wires,'' Phys. Rev. Lett. \textbf{80}, 5611 (1998), \href{https://arxiv.org/abs/cond-mat/9712269}{cond-mat/9712269}.
\bibitem{GT} K. Gaw\k{e}dzki and C. Tauber, ``Nonequilibrium transport through quantum-wire junctions and boundary defects for free massless bosonic fields,'' Nucl. Phys. B \textbf{896}, 138 (2015), \href{https://arxiv.org/abs/1501.07433}{arXiv:1501.07433}.
\bibitem{ChamonOA} C. Chamon, M. Oshikawa, and I. Affleck, ``Junctions of three quantum wires and the dissipative Hofstadter model,'' Phys. Rev. Lett. \textbf{91}, 206403 (2003), \href{https://arxiv.org/abs/cond-mat/0305121}{cond-mat/0305121}.
\bibitem{OCA2006} M. Oshikawa, C. Chamon, and I. Affleck, ``Junctions of three quantum wires,'' J. Stat. Mech. (2006) P02008, \href{https://arxiv.org/abs/cond-mat/0509675}{cond-mat/0509675}.
\bibitem{Rahmani} A. Rahmani, C.-Y. Hou, A. Feiguin, M. Oshikawa, C. Chamon, and I. Affleck, ``General method for calculating the universal conductance of strongly correlated junctions of multiple quantum wires,'' Phys. Rev. B \textbf{85}, 045120 (2012), \href{https://arxiv.org/abs/1108.4418}{arXiv:1108.4418}.
\bibitem{BernardDoyon2012} D. Bernard and B. Doyon, ``Energy flow in non-equilibrium conformal field theory,'' J. Phys. A \textbf{45}, 362001 (2012), \href{https://arxiv.org/abs/1202.0239}{arXiv:1202.0239}.
\bibitem{BernardDoyon2016} D. Bernard and B. Doyon, ``Conformal field theory out of equilibrium: a review,'' J. Stat. Mech. (2016) 064005, \href{https://arxiv.org/abs/1603.07765}{arXiv:1603.07765}.
\bibitem{Cardy1984} J. L. Cardy, ``Conformal invariance and surface critical behavior,'' Nucl. Phys. B \textbf{240}, 514 (1984).
\bibitem{Ishibashi} N. Ishibashi, ``The boundary and crosscap states in conformal field theories,'' Mod. Phys. Lett. A \textbf{4}, 251 (1989).
\bibitem{CardyLewellen} J. L. Cardy and D. C. Lewellen, ``Bulk and boundary operators in conformal field theory,'' Phys. Lett. B \textbf{259}, 274 (1991).
\bibitem{AffleckLudwigG} I. Affleck and A. W. W. Ludwig, ``Universal noninteger `ground-state degeneracy' in critical quantum systems,'' Phys. Rev. Lett. \textbf{67}, 161 (1991).
\bibitem{WongAffleck} E. Wong and I. Affleck, ``Tunneling in quantum wires: A boundary conformal field theory approach,'' Nucl. Phys. B \textbf{417}, 403 (1994), \href{https://arxiv.org/abs/cond-mat/9311040}{cond-mat/9311040}.
\bibitem{Oshikawa1996} M. Oshikawa and I. Affleck, ``Defect lines in the Ising model and boundary states on orbifolds,'' Phys. Rev. Lett. \textbf{77}, 2604 (1996), \href{https://arxiv.org/abs/hep-th/9606177}{hep-th/9606177}.
\bibitem{Oshikawa1997} M. Oshikawa and I. Affleck, ``Boundary conformal field theory approach to the critical two-dimensional Ising model with a defect line,'' Nucl. Phys. B \textbf{495}, 533 (1997), \href{https://arxiv.org/abs/cond-mat/9612187}{cond-mat/9612187}.
\bibitem{Bachas2002} C. Bachas, J. de Boer, R. Dijkgraaf, and H. Ooguri, ``Permeable conformal walls and holography,'' J. High Energy Phys. \textbf{06} (2002) 027, \href{https://arxiv.org/abs/hep-th/0111210}{hep-th/0111210}.
\bibitem{CardyBCFT} J. Cardy, ``Boundary conformal field theory,'' in \emph{Encyclopedia of Mathematical Physics} (Elsevier, Amsterdam, 2006), \href{https://arxiv.org/abs/hep-th/0411189}{hep-th/0411189}.
\bibitem{RecknagelSchomerusBook} A. Recknagel and V. Schomerus, \emph{Boundary Conformal Field Theory and the Worldsheet Approach to D-Branes} (Cambridge University Press, Cambridge, 2013).
\bibitem{AlekseevSchomerus} A. Yu. Alekseev and V. Schomerus, ``D-branes in the WZW model,'' Phys. Rev. D \textbf{60}, 061901 (1999), \href{https://arxiv.org/abs/hep-th/9812193}{hep-th/9812193}.
\bibitem{BPPZ} R. E. Behrend, P. A. Pearce, V. B. Petkova, and J.-B. Zuber, ``Boundary conditions in rational conformal field theories,'' Nucl. Phys. B \textbf{570}, 525 (2000), \href{https://arxiv.org/abs/hep-th/9908036}{hep-th/9908036}.
\bibitem{HofmanMaldacena} D. M. Hofman and J. Maldacena, ``Conformal collider physics: Energy and charge correlations,'' J. High Energy Phys. \textbf{05} (2008) 012, \href{https://arxiv.org/abs/0803.1467}{arXiv:0803.1467}.
\bibitem{FLPW} T. Faulkner, R. G. Leigh, O. Parrikar, and H. Wang, ``Modular Hamiltonians for deformed half-spaces and the averaged null energy condition,'' J. High Energy Phys. \textbf{09} (2016) 038, \href{https://arxiv.org/abs/1605.08072}{arXiv:1605.08072}.
\bibitem{HKT} T. Hartman, S. Kundu, and A. Tajdini, ``Averaged null energy condition from causality,'' J. High Energy Phys. \textbf{07} (2017) 066, \href{https://arxiv.org/abs/1610.05308}{arXiv:1610.05308}.
\bibitem{Oliviero2024} F. G. Oliviero, W. B. Fontana, and R. G. Pereira, ``Majorana Fermi surface state in a network of quantum spin chains,'' Phys. Rev. B \textbf{109}, 075135 (2024), \href{https://arxiv.org/abs/2310.09582}{arXiv:2310.09582}.
\bibitem{BachasIsing2013} C. Bachas, I. Brunner, and D. Roggenkamp, ``Fusion of critical defect lines in the 2D Ising model,'' J. Stat. Mech. (2013) P08008, \href{https://arxiv.org/abs/1303.3616}{arXiv:1303.3616}.
\bibitem{BachasBrunner2008} C. Bachas and I. Brunner, ``Fusion of conformal interfaces,'' J. High Energy Phys. \textbf{02} (2008) 085, \href{https://arxiv.org/abs/0712.0076}{arXiv:0712.0076}.
\bibitem{BachasWorldsheet2012} C. Bachas, I. Brunner, and D. Roggenkamp, ``A worldsheet extension of $O(d,d\mid\mathbb Z)$,'' J. High Energy Phys. \textbf{10} (2012) 039, \href{https://arxiv.org/abs/1205.4647}{arXiv:1205.4647}.
\bibitem{GKO1985} P. Goddard, A. Kent, and D. Olive, ``Virasoro algebras and coset space models,'' Phys. Lett. B \textbf{152}, 88 (1985).
\bibitem{GKO1986} P. Goddard, A. Kent, and D. Olive, ``Unitary representations of the Virasoro and super-Virasoro algebras,'' Commun. Math. Phys. \textbf{103}, 105 (1986).
\bibitem{QS} T. Quella and V. Schomerus, ``Symmetry breaking boundary states and defect lines,'' J. High Energy Phys. \textbf{06} (2002) 028, \href{https://arxiv.org/abs/hep-th/0203161}{hep-th/0203161}.
\bibitem{Cardy1989} J. L. Cardy, ``Boundary conditions, fusion rules and the Verlinde formula,'' Nucl. Phys. B \textbf{324}, 581 (1989).
\bibitem{FRS2002} J. Fuchs, I. Runkel, and C. Schweigert, ``TFT construction of RCFT correlators I: Partition functions,'' Nucl. Phys. B \textbf{646}, 353 (2002), \href{https://arxiv.org/abs/hep-th/0204148}{hep-th/0204148}.
\bibitem{PetkovaZuber} V. B. Petkova and J.-B. Zuber, ``Generalised twisted partition functions,'' Phys. Lett. B \textbf{504}, 157 (2001), \href{https://arxiv.org/abs/hep-th/0011021}{hep-th/0011021}.
\bibitem{GrahamWatts} K. Graham and G. M. T. Watts, ``Defect lines and boundary flows,'' J. High Energy Phys. \textbf{04} (2004) 019, \href{https://arxiv.org/abs/hep-th/0306167}{hep-th/0306167}.
\bibitem{BachasGaberdiel} C. Bachas and M. R. Gaberdiel, ``Loop operators and the Kondo problem,'' J. High Energy Phys. \textbf{11} (2004) 065, \href{https://arxiv.org/abs/hep-th/0411067}{hep-th/0411067}.
\bibitem{FFRS2007} J. Fr\"ohlich, J. Fuchs, I. Runkel, and C. Schweigert, ``Duality and defects in rational conformal field theory,'' Nucl. Phys. B \textbf{763}, 354 (2007), \href{https://arxiv.org/abs/hep-th/0607247}{hep-th/0607247}.
\bibitem{Verlinde1988} E. Verlinde, ``Fusion rules and modular transformations in 2D conformal field theory,'' Nucl. Phys. B \textbf{300}, 360 (1988).
\bibitem{Pendry} J. B. Pendry, ``Quantum limits to the flow of information and entropy,'' J. Phys. A \textbf{16}, 2161 (1983).
\bibitem{RegoKirczenow} L. G. C. Rego and G. Kirczenow, ``Quantized thermal conductance of dielectric quantum wires,'' Phys. Rev. Lett. \textbf{81}, 232 (1998), \href{https://arxiv.org/abs/cond-mat/9801238}{cond-mat/9801238}.
\bibitem{Schwab} K. Schwab, E. A. Henriksen, J. M. Worlock, and M. L. Roukes, ``Measurement of the quantum of thermal conductance,'' Nature \textbf{404}, 974 (2000).
\bibitem{Jezouin} S. Jezouin, F. D. Parmentier, A. Anthore, U. Gennser, A. Cavanna, Y. Jin, and F. Pierre, ``Quantum limit of heat flow across a single electronic channel,'' Science \textbf{342}, 601 (2013).
\bibitem{Safi} I. Safi and H. J. Schulz, ``Transport in an inhomogeneous interacting one-dimensional system,'' Phys. Rev. B \textbf{52}, R17040 (1995), \href{https://arxiv.org/abs/cond-mat/9505079}{cond-mat/9505079}.
\bibitem{MaslovStone} D. L. Maslov and M. Stone, ``Landauer conductance of Luttinger liquids with leads,'' Phys. Rev. B \textbf{52}, R5539 (1995), \href{https://arxiv.org/abs/cond-mat/9505098}{cond-mat/9505098}.
\bibitem{Ponomarenko} V. V. Ponomarenko, ``Renormalization of the one-dimensional conductance in the Luttinger-liquid model,'' Phys. Rev. B \textbf{52}, R8666 (1995), \href{https://arxiv.org/abs/cond-mat/9507089}{cond-mat/9507089}.
\bibitem{Hou} C.-Y. Hou, A. Rahmani, A. E. Feiguin, and C. Chamon, ``Junctions of multiple quantum wires with different Luttinger parameters,'' Phys. Rev. B \textbf{86}, 075451 (2012), \href{https://arxiv.org/abs/1205.2125}{arXiv:1205.2125}.
\bibitem{LevitovLesovik} L. S. Levitov and G. B. Lesovik, ``Charge distribution in quantum shot noise,'' JETP Lett. \textbf{58}, 230 (1993).
\bibitem{LLL} L. S. Levitov, H. Lee, and G. B. Lesovik, ``Electron counting statistics and coherent states of electric current,'' J. Math. Phys. \textbf{37}, 4845 (1996), \href{https://arxiv.org/abs/cond-mat/9607137}{cond-mat/9607137}.
\bibitem{Schoenhammer} K. Sch\"onhammer, ``Full counting statistics for noninteracting fermions: Exact results and the Levitov-Lesovik formula,'' Phys. Rev. B \textbf{75}, 205329 (2007), \href{https://arxiv.org/abs/cond-mat/0701620}{cond-mat/0701620}.
\bibitem{Klich} I. Klich and L. Levitov, ``Quantum noise as an entanglement meter,'' Phys. Rev. Lett. \textbf{102}, 100502 (2009), \href{https://arxiv.org/abs/0804.1377}{arXiv:0804.1377}.
\bibitem{ChoiEtAl} Y. Choi, H. Ha, D. Kim, Y. Kusuki, S. Ohyama, and S. Ryu, ``Higher structures on boundary conformal manifolds: Higher Berry phase and boundary conformal field theory,'' Phys. Rev. D \textbf{113}, 106005 (2026), \href{https://arxiv.org/abs/2507.12525}{arXiv:2507.12525}.
\bibitem{AffleckBCC} I. Affleck, ``Boundary condition changing operations in conformal field theory and condensed matter physics,'' Nucl. Phys. B Proc. Suppl. \textbf{58}, 35 (1997), \href{https://arxiv.org/abs/hep-th/9611064}{hep-th/9611064}.
\bibitem{GKSW} D. Gaiotto, A. Kapustin, N. Seiberg, and B. Willett, ``Generalized global symmetries,'' J. High Energy Phys. \textbf{02} (2015) 172, \href{https://arxiv.org/abs/1412.5148}{arXiv:1412.5148}.
\bibitem{CLSWY} C.-M. Chang, Y.-H. Lin, S.-H. Shao, Y. Wang, and X. Yin, ``Topological defect lines and renormalization group flows in two dimensions,'' J. High Energy Phys. \textbf{01} (2019) 026, \href{https://arxiv.org/abs/1802.04445}{arXiv:1802.04445}.
\bibitem{AffleckHaldane} I. Affleck and F. D. M. Haldane, ``Critical theory of quantum spin chains,'' Phys. Rev. B \textbf{36}, 5291 (1987).
\bibitem{ALKondo1991a} I. Affleck and A. W. W. Ludwig, ``The Kondo effect, conformal field theory and fusion rules,'' Nucl. Phys. B \textbf{352}, 849 (1991).
\bibitem{ALKondo1991b} I. Affleck and A. W. W. Ludwig, ``Critical theory of overscreened Kondo fixed points,'' Nucl. Phys. B \textbf{360}, 641 (1991).
\end{thebibliography}
\end{document}